\documentclass[twocolumn,prd,amsmath,amssymb,showkeys,showpacs,superscriptaddress,floatfix,footinbib]{revtex4}

\usepackage{graphicx}
\usepackage{optional}
\graphicspath{{Img/}}

\def\units#1{\hbox{$\,{\rm #1}$}}
\def\degrees{\hbox{$^\circ$}}

\begin{document}



\title{Constraints on dark matter models from a \emph{Fermi} LAT search for high-energy cosmic-ray electrons from the Sun}

\author{M.~Ajello}
\affiliation{W. W. Hansen Experimental Physics Laboratory, Kavli Institute for Particle Astrophysics and Cosmology, Department of Physics and SLAC National Accelerator Laboratory, Stanford University, Stanford, CA 94305, USA}
\author{W.~B.~Atwood}
\affiliation{Santa Cruz Institute for Particle Physics, Department of Physics and Department of Astronomy and Astrophysics, University of California at Santa Cruz, Santa Cruz, CA 95064, USA}
\author{L.~Baldini}
\affiliation{Istituto Nazionale di Fisica Nucleare, Sezione di Pisa, I-56127 Pisa, Italy}
\author{G.~Barbiellini}
\affiliation{Istituto Nazionale di Fisica Nucleare, Sezione di Trieste, I-34127 Trieste, Italy}
\affiliation{Dipartimento di Fisica, Universit\`a di Trieste, I-34127 Trieste, Italy}
\author{D.~Bastieri}
\affiliation{Istituto Nazionale di Fisica Nucleare, Sezione di Padova, I-35131 Padova, Italy}
\affiliation{Dipartimento di Fisica ``G. Galilei", Universit\`a di Padova, I-35131 Padova, Italy}
\author{R.~Bellazzini}
\affiliation{Istituto Nazionale di Fisica Nucleare, Sezione di Pisa, I-56127 Pisa, Italy}
\author{B.~Berenji}
\affiliation{W. W. Hansen Experimental Physics Laboratory, Kavli Institute for Particle Astrophysics and Cosmology, Department of Physics and SLAC National Accelerator Laboratory, Stanford University, Stanford, CA 94305, USA}
\author{R.~D.~Blandford}
\affiliation{W. W. Hansen Experimental Physics Laboratory, Kavli Institute for Particle Astrophysics and Cosmology, Department of Physics and SLAC National Accelerator Laboratory, Stanford University, Stanford, CA 94305, USA}
\author{E.~D.~Bloom}
\affiliation{W. W. Hansen Experimental Physics Laboratory, Kavli Institute for Particle Astrophysics and Cosmology, Department of Physics and SLAC National Accelerator Laboratory, Stanford University, Stanford, CA 94305, USA}
\author{E.~Bonamente}
\affiliation{Istituto Nazionale di Fisica Nucleare, Sezione di Perugia, I-06123 Perugia, Italy}
\affiliation{Dipartimento di Fisica, Universit\`a degli Studi di Perugia, I-06123 Perugia, Italy}
% ...

\date{\today}

\begin{abstract}

During its first year of data taking, the Large Area Telescope
(LAT) onboard the {\em Fermi} Gamma-Ray Space Telescope has 
collected a large sample of high-energy cosmic-ray electrons and 
positrons (CREs). We present the results of a directional analysis of the CRE events, 
in which we searched for a flux excess correlated with the direction of the Sun. Two different
and complementary analysis approaches were implemented, 
and neither yielded evidence of a significant CRE
flux excess from the Sun.  We derive upper limits on the CRE flux 
from the Sun's direction, and use these bounds to constrain
two classes of dark matter models which predict a solar CRE flux:
(1) models in which dark matter annihilates to CREs via a light intermediate
state, and (2) inelastic dark matter models in which dark matter annihilates to CREs. 

\end{abstract}


\maketitle

\section{Introduction}

In the last decades the searches for a dark matter (DM) signal from 
the Sun were performed looking for possible excesses of neutrinos 
or gamma-rays associated with the Sun's direction. However, as it 
was noted in Ref.~\cite{Schuster:2009fc}, 
several DM models that have been recently developed to explain various
experimental results also imply an associated solar flux of
high-energy cosmic-ray electrons and positrons (CREs). On 
the other hand, no known
astrophysical mechanisms are expected to generate a significant 
high-energy CRE $(>100 \units{GeV})$
excess associated with the Sun.

A class of models in which DM annihilates to CREs through a new light intermediate 
state $\phi$ \cite{Pospelov:2007mp,ArkaniHamed:2008qn} has 
been considered to explain the excesses in local CRE fluxes reported 
by PAMELA~\cite{Adriani:2008zr},
ATIC~\cite{:2008zzr}, and \emph{Fermi}~\cite{Abdo:2009zk,Ackermann:2010ij}.
In these scenarios DM particles captured by the Sun through elastic scattering interactions
would annihilate to $\phi$ pairs in the Sun's core, and if the $\phi$ could escape
the surface of the Sun before decaying to CREs, these models could
produce an observable CRE flux.

Another class of models in which DM scatters off of nucleons
predominantly via inelastic scattering 
has been proposed as a means of reconciling the results of 
DAMA and DAMA/LIBRA~\cite{Bernabei:2008yi,Bernabei:2010mq} 
with CDMS-II~\cite{Ahmed:2009zw,Ahmed:2010hw} and
other experiments (e.g., \cite{Chang:2008gd,Finkbeiner:2009ug}; see
also \cite{Savage:2008er} for a comprehensive discussion of experimental
constraints).  If DM is captured
by the Sun only through inelastic scattering (iDM), this could lead to
a non-negligible fraction of DM annihilating outside of the
Sun's surface.  For models in which iDM annihilates to CREs, an observable
flux at energies above a few tens of GeV could be produced. 

During its first year of operation, the Large Area Telescope (LAT) 
onboard the {\em Fermi} satellite~\cite{Atwood:2009ez} has collected a 
substantial number of CRE events, which has allowed a precise measurement 
of the energy spectrum over a broad
energy range from a few $\units{GeV}$ 
up to $1\units{TeV}$~\cite{Abdo:2009zk,Ackermann:2010ij}. 
Furthermore, a directional analysis of the high-energy CRE
events was performed in the Galactic reference 
frame~\cite{Ackermann:2010ip}, and showed no evidence of anisotropies. 

In this paper we use the high-energy CRE data set to search 
for flux variations correlated with the Sun's direction. 
Since the Sun is moving with respect to the Galactic reference frame, 
the previously-reported absence of anisotropies in the CRE flux observed in the Galactic 
frame does not necessarily imply a negative result.

\section{Data selection} 
\label{sec:datasel}

The {\em Fermi} LAT is a pair-conversion telescope designed
to detect gamma rays in the energy range from $20 \units{MeV}$
to more than $300 \units{GeV}$. A full description of the 
apparatus is given in~\cite{Atwood:2009ez}. Even though it is 
a photon detector, it has been demonstrated that the LAT 
is also an excellent CRE 
detector~\cite{Abdo:2009zk,Ackermann:2010ij,Ackermann:2010ip}.
For this analysis we used the CRE data sample collected 
by the LAT during its first year of operation, starting from 
August 4, 2008. The event selection was performed in the 
same way as in Ref.~\cite{Ackermann:2010ip};
approximately $1.35\times10^{6}$ CRE events with energies 
larger than $60\units{GeV}$ passed the selection cuts.
As discussed in Ref.~\cite{Ackermann:2010ip}, the energy threshold
of $60\units{GeV}$ was chosen  
because it is higher than the geomagnetic cutoff 
in any part of  {\em Fermi}'s orbit.

Unlike gamma rays, CREs are deflected by interactions 
with magnetic fields encountered during their propagation in
interstellar space. 
In particular, CREs coming from the Sun are
deflected by both the Sun's and the Earth's
magnetic fields.

Geomagnetic effects on CREs have been studied
using a code that reconstructs the trajectories of 
charged particles in the Earth's magnetic field 
based on the International Geomagnetic Reference Field (IGRF) 
model~\cite{IGRF}.  
Since the {\em Fermi} LAT cannot measure the sign of the electric charge, 
we associated both an electron and a positron track with 
each CRE event detected by the LAT.
Each track starts from the 
detection point with the same energy of the event and with 
a direction opposite to that of the event, and ends at a 
very large distance (larger than 100 Earth-radii from 
the Earth's center). 


The distribution of deflection angles at different energies was
analyzed. The simulation demonstrated that, at energies 
above $20\units{GeV}$,  $90\%$ of the particles are deflected 
with respect to the original direction within an angle 
$\delta_{90 \% }$ given by the approximate formula:  

\begin{equation}
\delta_{90\% } \approx \frac{2.8\degrees}{E(\units{TeV})} 
\label{eq:deflection}
\end{equation} 
where $E$ is the particle energy. 
Hence, due to the geomagnetic field, the reconstructed directions 
of CREs with energies above $100\units{GeV}$ detected by the LAT and 
coming from any given direction of the sky will be spread over 
a cone with an angular radius of about $30\degrees$ centered on the 
original incoming direction. 
  
The directions of incoming CREs are also affected by the
Heliospheric Magnetic Field. A detailed study of its effects 
on CREs is beyond the scope of this work, 
however in Ref.~\cite{Roberts:2010yh},
it was shown that CREs with energies of
several hundreds of $\units{GeV}$ can travel through the center 
of the solar system without experiencing significant 
deflections.

CREs travelling in the Solar System may also suffer energy losses, mainly due to the Inverse Compton (IC) scatterings on the
photons emitted by the Sun and to the synchrotron radiation (SR) emitted in the interactions with the Heliospheric Magnetic
Field. To study the energy loss processes of CREs travelling from the Sun to the Earth we implemented a simple toy model, in
which we assumed that CREs propagate from the Sun's surface to the Earth in straight lines and with velocity $c$.
Following Ref.~\cite{Orlando:2008uk}, we assumed that the Sun can be modeled as a black body with a temperature 
of $5777\units{K}$ and with a photon density given by:

\begin{equation}
\label{eq:blackbody}
N_{ph}(\epsilon,r) = 0.5 n_{bb}(\epsilon) \left[ 1 - \sqrt{ 1 - \frac{R_{\odot}^{2}}{r^{2}}  } ~ \right]
\end{equation}
where $n_{bb}(\epsilon)$ is the blackbody photon energy density (Planck's equation), $R_{\odot}$ is the solar radius and $r$ is the
distance from the center of the Sun. The IC energy loss rate of CREs was then evaluated as in Ref.~\cite{Schlickeiser:2009qq} as:

\begin{equation}
\label{eq:ICloss}
- \left( \frac{dE}{dt} \right)_{IC} = \frac{4}{3} \sigma_{T} c W \beta^{2}
\frac{\gamma_{k}^{2} \gamma^{2}}{\gamma_{k}^{2}  + \gamma^{2}}
\end{equation}
where  $\beta c$ and $\gamma$ are respectively the velocity and the Lorentz factor of the CRE, $W$ is the photon energy
density evaluated from eq.~\ref{eq:blackbody}, $\sigma_{T}$ is the Thomson cross section and $\gamma_{k}$ is given by:

\begin{equation}
\gamma_{k} = \frac{3 \sqrt{5} m_{e} c^{2}}{8 \pi k_{B} T}
\end{equation}
where $m_{e}$ is the electron mass, $k_{B}$ is the Boltzmann's constant and $T=5777K$ is the temperature of the Sun's surface.
The evaluation of the SR energy loss rate is not easy, because the structure of the Heliospheric Magnetic Field is rather
complex~\cite{Parker:1958zz}. However, as a first approximation, we assumed that the strength of the Heliospheric Magnetic Field drops from the Sun's surface as:

\begin{equation}
B(r) = B_{0} \frac{R_{\odot}^{2}}{r^{2}}
\end{equation}
where $B_{0}=1\units{gauss}$ is the strength of the field on the Sun's surface.
We did not include the contribution of the Geomagnetic field to synchrotron energy losses because, 
even though the field strength at the LAT altitude is of the order of $1\units{gauss}$, 
the path length of CREs in the Geomagnetic field is of the order of a few Earth radii, 
which is negligible with respect to the path length in the Heliospheric Magnetic field, which is
of the order of a few solar radii.

The synchrotron energy loss rate of CREs was then calculated as~\cite{Rybicki}:

\begin{equation}
\label{eq:syncloss}
- \left( \frac{dE}{dt} \right)_{S} = \frac{4}{3} \sigma_{T} c W_{B} \beta^{2} \gamma^{2}
\end{equation}
where $W_{B}$ is the magnetic field energy density that includes only the contribution from
the Heliospheric Magnetic Field and, in our model, results to be negliglible with respect to the IC energy
loss rate.
Using eqs.~\ref{eq:ICloss} and~\ref{eq:syncloss}, we calculated that CREs in the energy range from 
$60\units{GeV}$ to $1\units{TeV}$ travelling from the Sun to the Earth lose 
no more than $2\%$ of their initial energy. 
Therefore, in the calculations of the following sections, we will neglect all energy loss processes.

\section{Data analysis and results}
 
To study the CRE flux from the Sun's direction and to search for
variations with respect to the average flux, we implemented
two complementary analysis approaches: (i) flux asymmetry 
analysis and (ii) comparison of the solar
flux with the isotropic flux.

\subsection{Flux asymmetry studies}
\label{sec:fluxasym}

\begin{figure*}[!ht]
\includegraphics[width=0.48\linewidth]{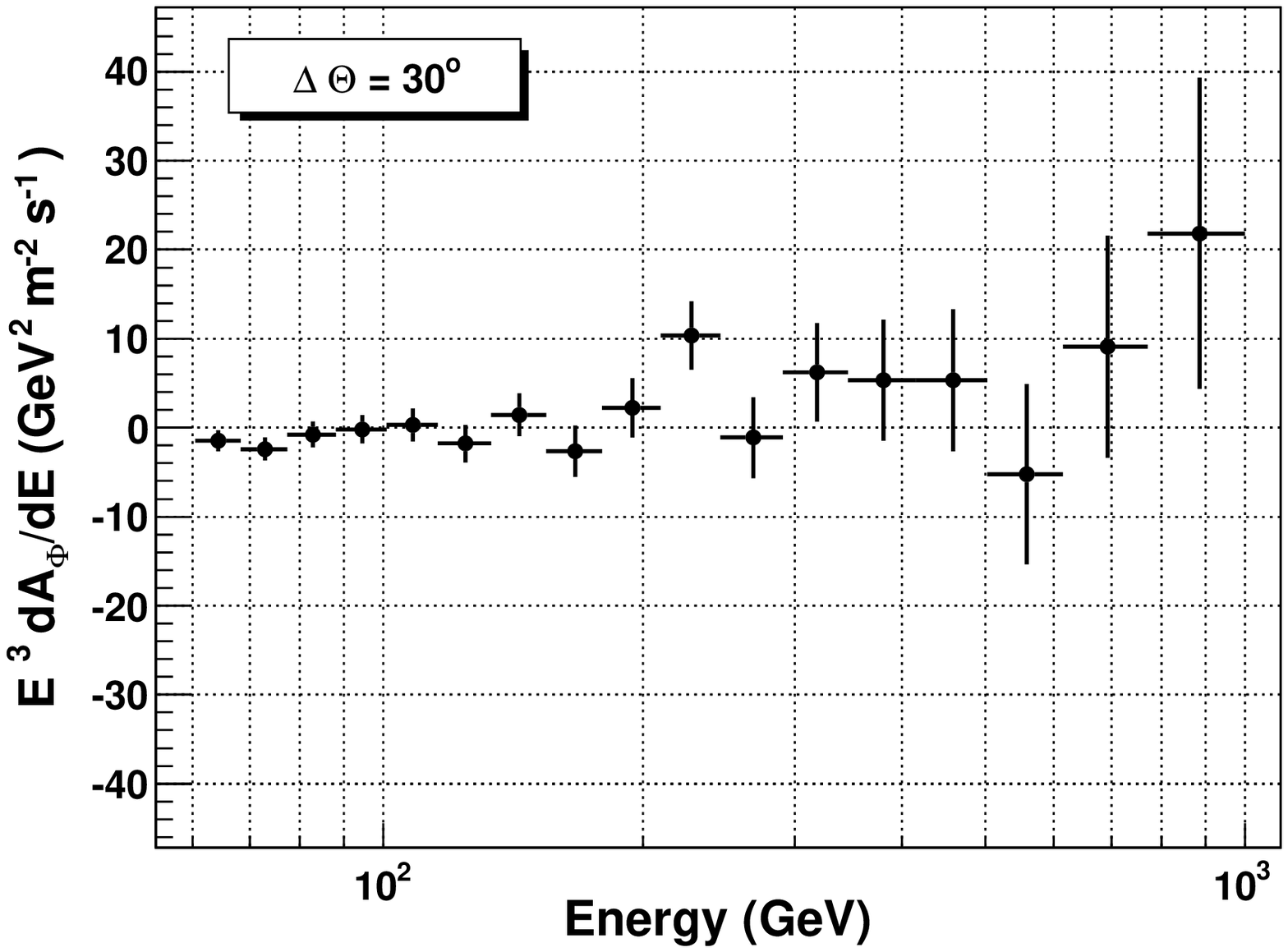}
\includegraphics[width=0.48\linewidth]{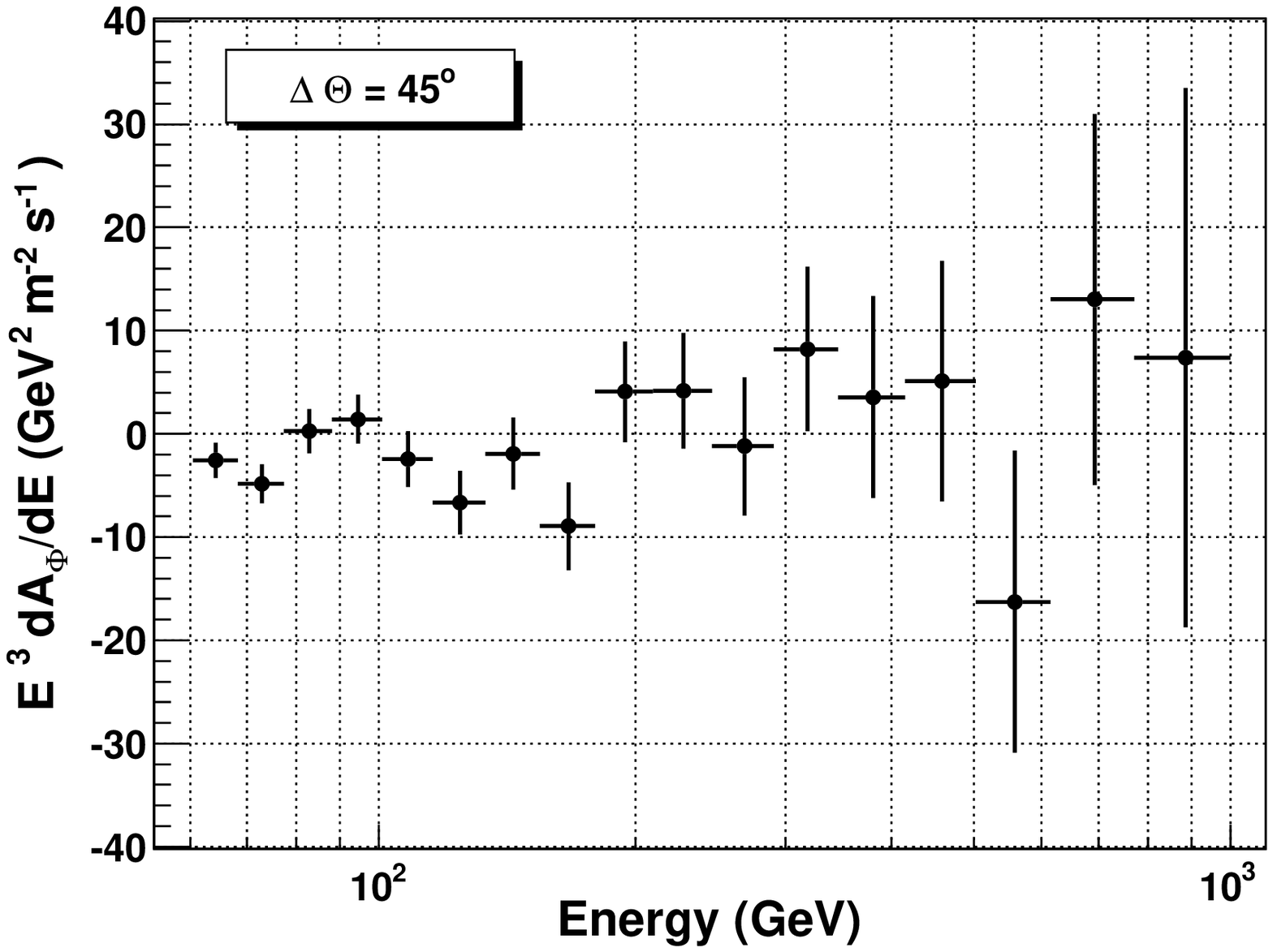}
\includegraphics[width=0.48\linewidth]{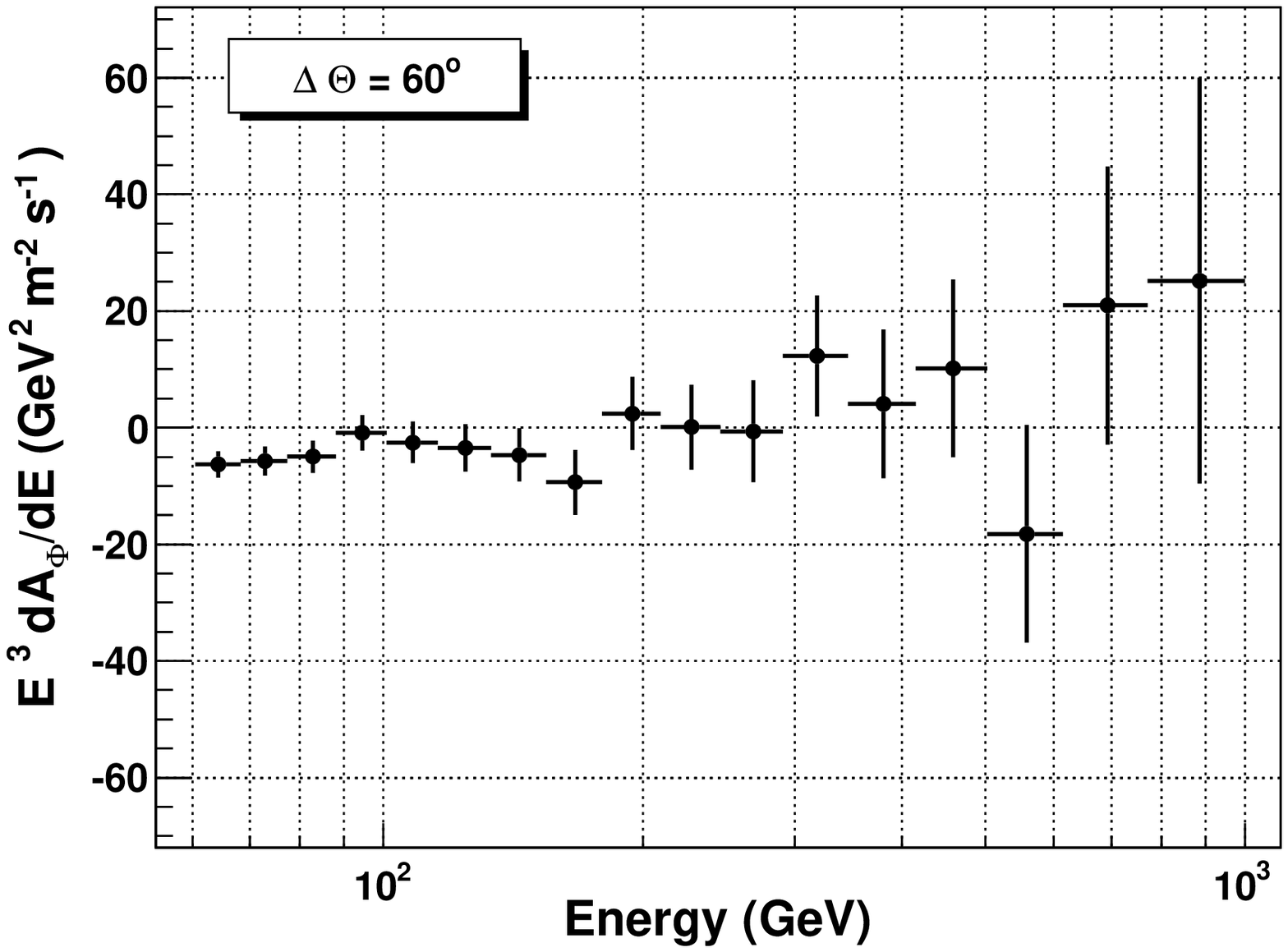}
\includegraphics[width=0.48\linewidth]{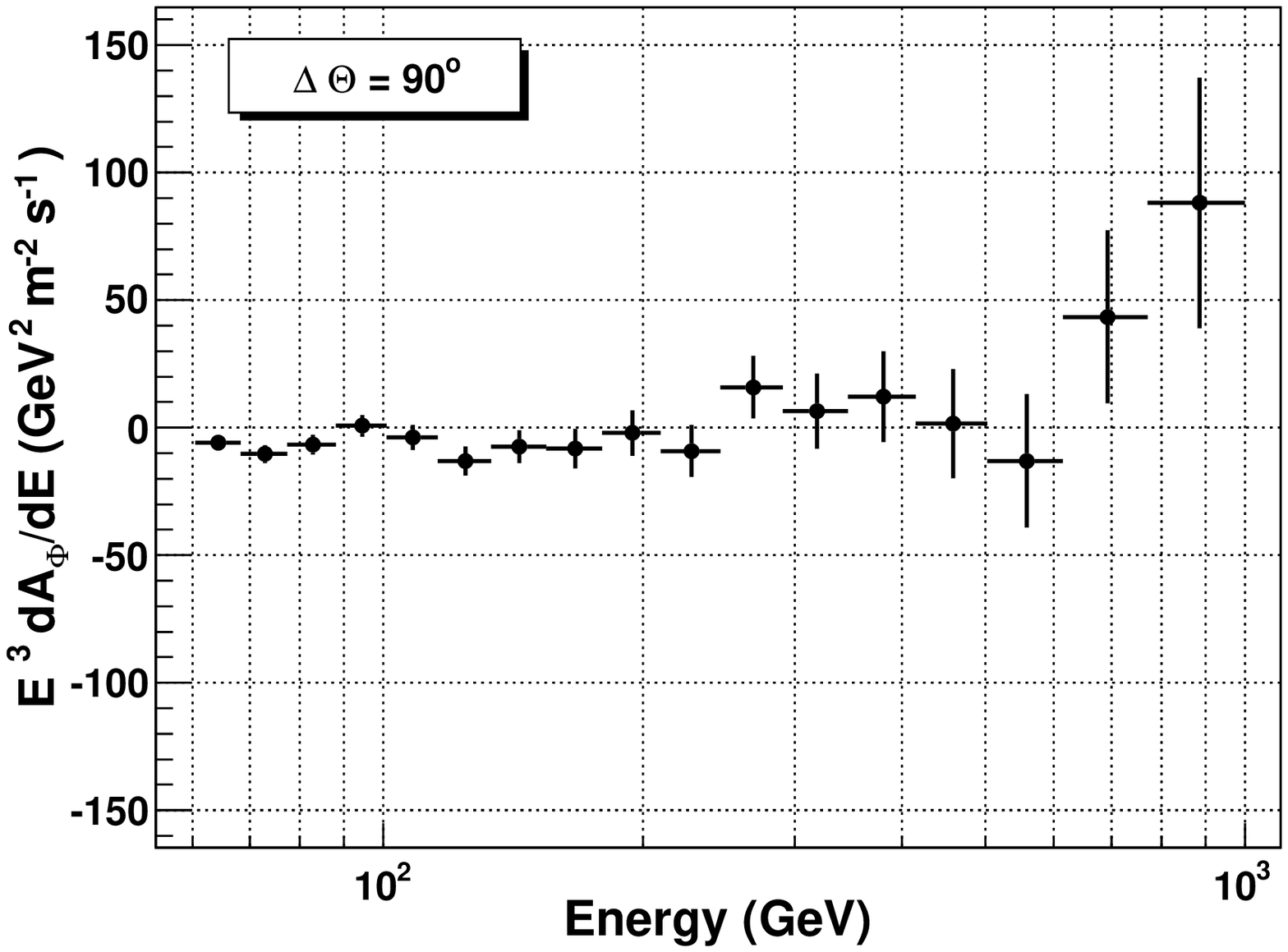}
\caption{Differential flux asymmetry between real and fake Sun evaluated
in cones with angular radii $\Delta \Theta = 30\degrees$ 
(top left panel), $45\degrees$ (top right panel), 
$60 \degrees$ (bottom left panel) and $90\degrees$ 
(bottom right panel). The fluxes are multiplied by
$E^{3}$ (the energy values correspond to the bin centers) 
since the energy spectrum of CREs is
approximately proportional to $E^{-3}$ in this energy range.  
Only statistical error bars are shown.
\label{fig:e3flux}}
\end{figure*}

\begin{table*}[!ht]
\begin{tabular}{||c||c|c|c||}
\hline
Angular radius & Maximum deviation ($\sigma_{max}$) & $P(|\sigma_{max}|)$ & 
$P(|\sigma|>|\sigma_{max}|)$ \\
\hline
$30 \degrees$ &  2.690 & 0.007 & 0.113 \\
\hline
$45 \degrees$ & -2.542 & 0.011 & 0.171 \\
\hline
$60 \degrees$ & -2.806 & 0.005 & 0.082 \\
\hline
$90 \degrees$ & -2.947 & 0.003 & 0.050 \\
\hline
\end{tabular}
\caption{For each cone used for the flux asymmetry analysis 
the maximum deviations (either positive or negative) from the null 
value are shown and the corresponding probabilities of observing
larger values in the hypothesis of null flux asymmetry. 
The last column shows the probability of finding at least one energy
bin with a larger flux asymmetry than the maximum observed value. 
\label{tab:prob1}}
\end{table*}

This approach compares the CRE flux from the 
Sun with the flux from a fake source (fake Sun) placed in the
sky position opposite to that of the Sun. 
To perform our analyses, we chose a custom reference frame
derived from ecliptic coordinates. The ecliptic coordinates
associated with each CRE event were evaluated from equatorial
coordinates using the formulae in Ref.~\cite{Duffett}.
Indicating with $(\lambda, \beta)$ the pair of ecliptic coordinates 
(longitude and latitude, respectively) associated with any given 
direction, the Sun's direction will always lie in the plane $\beta=0$.
In fact, since the Sun is moving eastwards along the path 
of the ecliptic, its ecliptic latitude will always be zero by
definition, while its ecliptic longitude will always increase,
describing a complete $360\degrees$ cycle in one year~\cite{Duffett}.
In our custom reference frame, the coordinates associated with
each direction are defined as:

\begin{equation}
\left\{
\begin{array}{l}
\lambda^{\prime} = \lambda - \lambda_{Sun} \\
\beta^{\prime} = \beta 	
\end{array}
\right.
\label{eq:coord}
\end{equation}
where $\lambda_{Sun}$ is the ecliptic longitude of the Sun,
evaluated from the Sun's ephemeris using a software
interfaced to the 
JPL libraries~\cite{JPL}.
In this reference frame, the Sun's coordinates will always be 
$(\lambda^{\prime}_{Sun}=0\degrees, \beta^{\prime}_{Sun}=0\degrees)$.
On the other hand, the fake Sun will always be located 
at the coordinates 
$(\lambda^{\prime}_{fake~Sun}=180\degrees, 
\beta^{\prime}_{fake~Sun}=0\degrees)$.

Due to the geomagnetic field's effects
on CRE trajectories described in 
\S\ref{sec:datasel}, we consider the fluxes from 
extended sky regions centered
on the Sun (and on the fake Sun). In particular,
we compare the CRE fluxes from directions 
within cones of angular radii $\Delta \Theta$, 
centered on the position of the Sun and the fake Sun. 
According to Eq.~\ref{eq:deflection}, $90\%$ of 
CREs with energies of $100 \units{GeV}$ are deflected
within a cone of about $30\degrees$ angular radius,
and so
we chose this value as the minimum angular 
radius of the sky regions to be investigated
because the DM models discussed in Ref.~\cite{Schuster:2009fc}
predict a CRE flux excess from the Sun 
in the energy range above $100\units{GeV}$.

To measure the fluxes from different sky regions, we first 
divided the sky into a grid of pixels, then evaluated 
the CRE fluxes from individual pixels (each pixel was treated
as a point source), and finally integrated the 
fluxes from the pixels belonging to the selected sky regions. 
We used the HEALPix~\cite{Gorski:2004by} pixelization
scheme, and divided the sky into 
$12288$ equal-area pixels, each covering a solid angle of about $10^{-3} \units{sr}$.
The CRE differential fluxes from individual pixels were evaluated 
according to the following equation:

\begin{equation}
\cfrac{d\Phi_{i} (E)}{dE} = \frac{1}{\Delta E}
\cfrac{N_{i}(E) \times (1-c(E))}
{\mathcal{E}_{i}(E)}
\label{eq:fluxdiff}
\end{equation} 
where $d\Phi_{i}(E)/dE$ is the differential CRE flux 
(expressed in particles per unit energy, 
unit area and unit time) in the energy interval 
$[E, E+\Delta E]$ from the $i$th pixel,
$N_{i}(E)$ is the number of observed CRE events from the 
$i$th pixel with energies between $E$ and $E+\Delta E$, 
$c(E)$ is the residual contamination (the contamination
values are reported in Ref.~\cite{Ackermann:2010ij})
and $\mathcal{E}_{i}(E)$ is the exposure of the $i$th
pixel, which is calculated taking into account the 
effective area of the instrument, and the live time
of the $i$th pixel. The dependence of the effective area 
on the CRE direction in the instrument, expressed in terms 
of the off-axis and azimuth angles $\theta$ and $\phi$, 
is also taken into account in the calculation.

The CRE flux from a cone of angular radius $\Delta \Theta$ centered 
on the Sun is then given by:

\begin{equation}
\cfrac{d \Phi_{Sun} (E | \Delta \Theta)}{dE} = 
\sum_{i \in  ROI(\Delta \Theta)} \cfrac{d \Phi_{i} (E)}{dE}   
\label{eq:fluxroi}
\end{equation}
where $ROI (\Delta \Theta)$ denotes the set of pixels
(region of interest) at an angular distance less than $\Delta \Theta$ from 
the Sun. The flux from the fake Sun is evaluated in
a similar way. The flux asymmetry can then be evaluated as: 

\begin{equation}
\cfrac{dA_{\Phi}(E|\Delta \Theta)}{dE} = \cfrac{d\Phi_{Sun}(E | \Delta \Theta)}{dE} 
- \cfrac{d\Phi_{Fake~Sun}(E | \Delta \Theta)}{dE}
\label{eq:asymmetry1}
\end{equation}
The variable $dA_{\Phi}(E|\Delta \Theta)/dE$ defined in 
Eq.~\ref{eq:asymmetry1} is the difference between
the CRE flux from the Sun 
and the fake Sun; the flux of the fake Sun is
assumed to be representative of the average CRE flux
across the sky.  
Positive (negative) values of $dA_{\Phi}(E | \Delta \Theta)/dE$  
indicate an excess (deficit) of CREs from the Sun.
We emphasize that this approach relies on the assumption that 
the flux from the fake Sun region is representative of the 
average CRE flux. 

In Fig.~\ref{fig:e3flux} the differential CRE flux asymmetries 
$dA_{\Phi}(E|\Delta \Theta)/dE$ between the real and the fake Sun
are shown for four different ROIs, with angular radii 
of $30\degrees$, $45\degrees$, $60 \degrees$ and $90\degrees$.
No significant CRE flux excesses or deficits 
from the Sun are observed at any energy. 
In the plots of Fig.~\ref{fig:e3flux} only statistical 
error bars are shown. As pointed out in
Ref.~\cite{Ackermann:2010ij}, the main source of 
systematic uncertainties in the evaluation of  
CRE fluxes is the imperfect knowledge of the 
detector's effective area. Assuming that it 
is affected only by a normalization 
error, when calculating the error on the 
flux differences $dA_{\Phi}(E|\Delta \Theta)/dE$, 
the contribution from the normalization error 
will be proportional to $|dA_{\Phi}(E|\Delta \Theta)/dE|$
(see the discussion in Ref.~\cite{D'Agostini:1993uj}), 
and therefore it will be negligible with respect to the statistical error.

Assuming that the measured flux asymmetries in each 
energy bin behave as Gaussian random variables, we 
expressed the excesses and deficits (with respect to 
the hypothesis of a null flux asymmetry) in units of $\sigma$
($\sigma$ is the statistical error associated with each measurement),
and evaluated the corresponding probabilities of measuring larger 
excesses or deficits assuming the null hypothesis. 
Table~\ref{tab:prob1} shows, for each value of the angular 
radius $\Delta \Theta$, the maximum observed deviations from the null 
flux asymmetry in units of $\sigma$, and the corresponding probabilities 
of measuring larger flux asymmetries in the null hypothesis. 
As shown in Table~\ref{tab:prob1}, the flux asymmetries
in all of the ROIs are always within $3 \sigma$ 
of zero. The last column of Table~\ref{tab:prob1} shows
the probabilities of finding, in each ROI, 
at least one energy bin with a flux asymmetry larger than
the maximum observed value. The probabilities
were calculated assuming that the flux asymmetries
measured in each of the $17$ energy bins used in our 
analysis are uncorrelated. 
The calculations were performed taking only
statistical errors into account; if systematic errors
were also taken into account, the significance of
the deviations of the flux asymmetries from zero 
would be smaller.

The same analysis was repeated for integral fluxes
above various energy thresholds
and again no evidence of flux asymmetries was found.

\subsubsection{Evaluation of statistical upper limits on the CRE flux asymmetry}

\begin{figure*}
\includegraphics[width=0.48\linewidth]{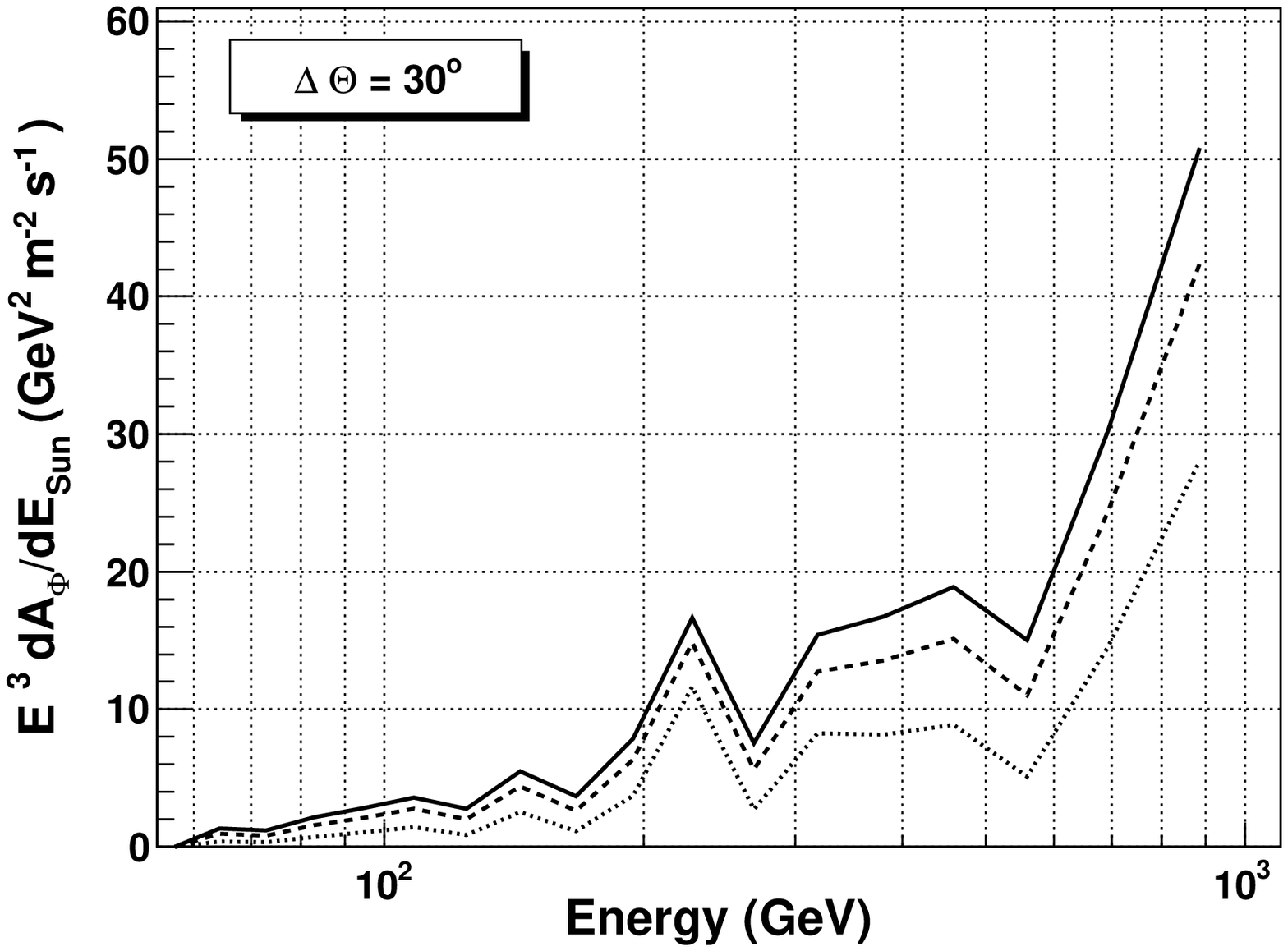}
\includegraphics[width=0.48\linewidth]{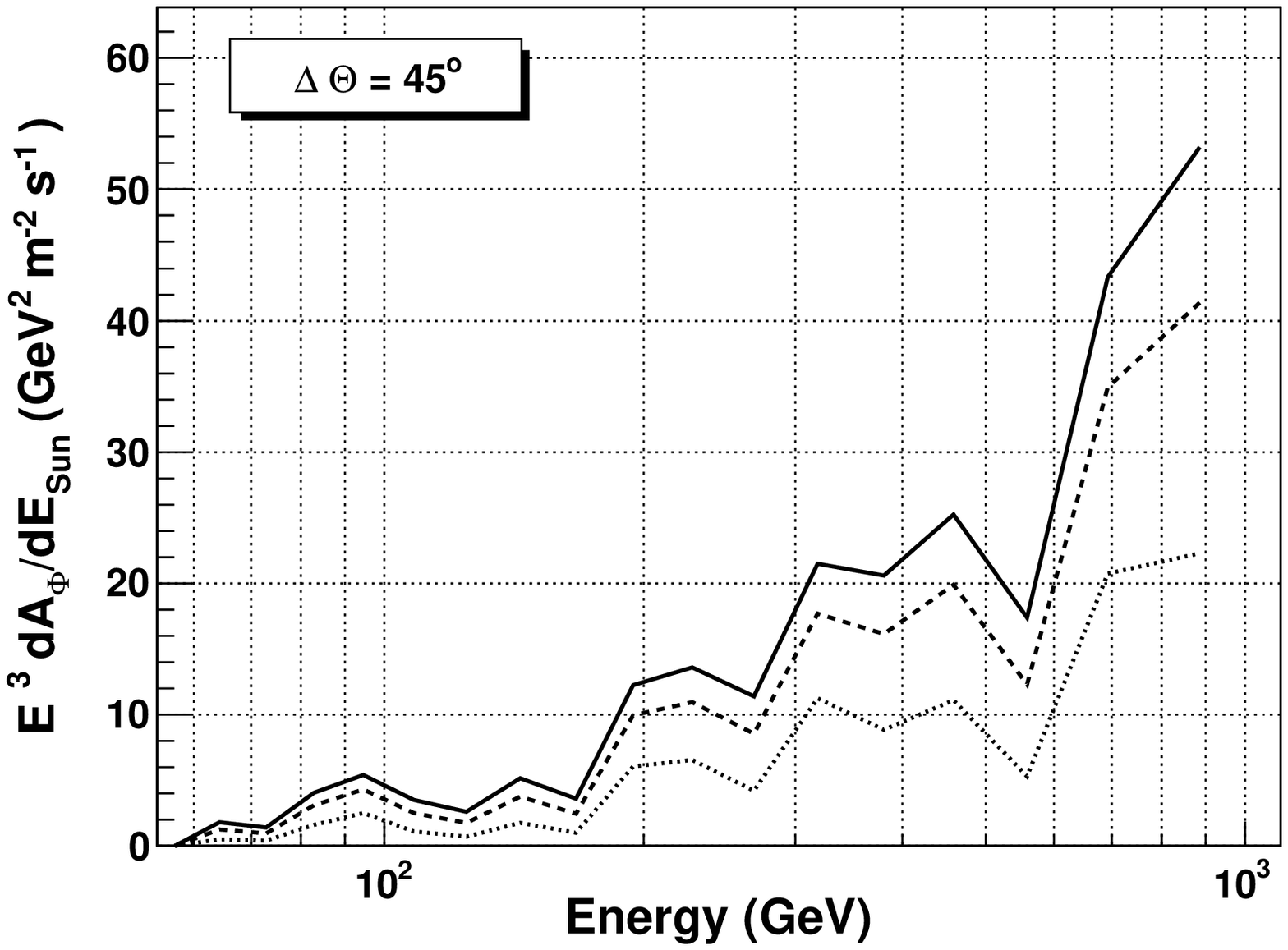}
\includegraphics[width=0.48\linewidth]{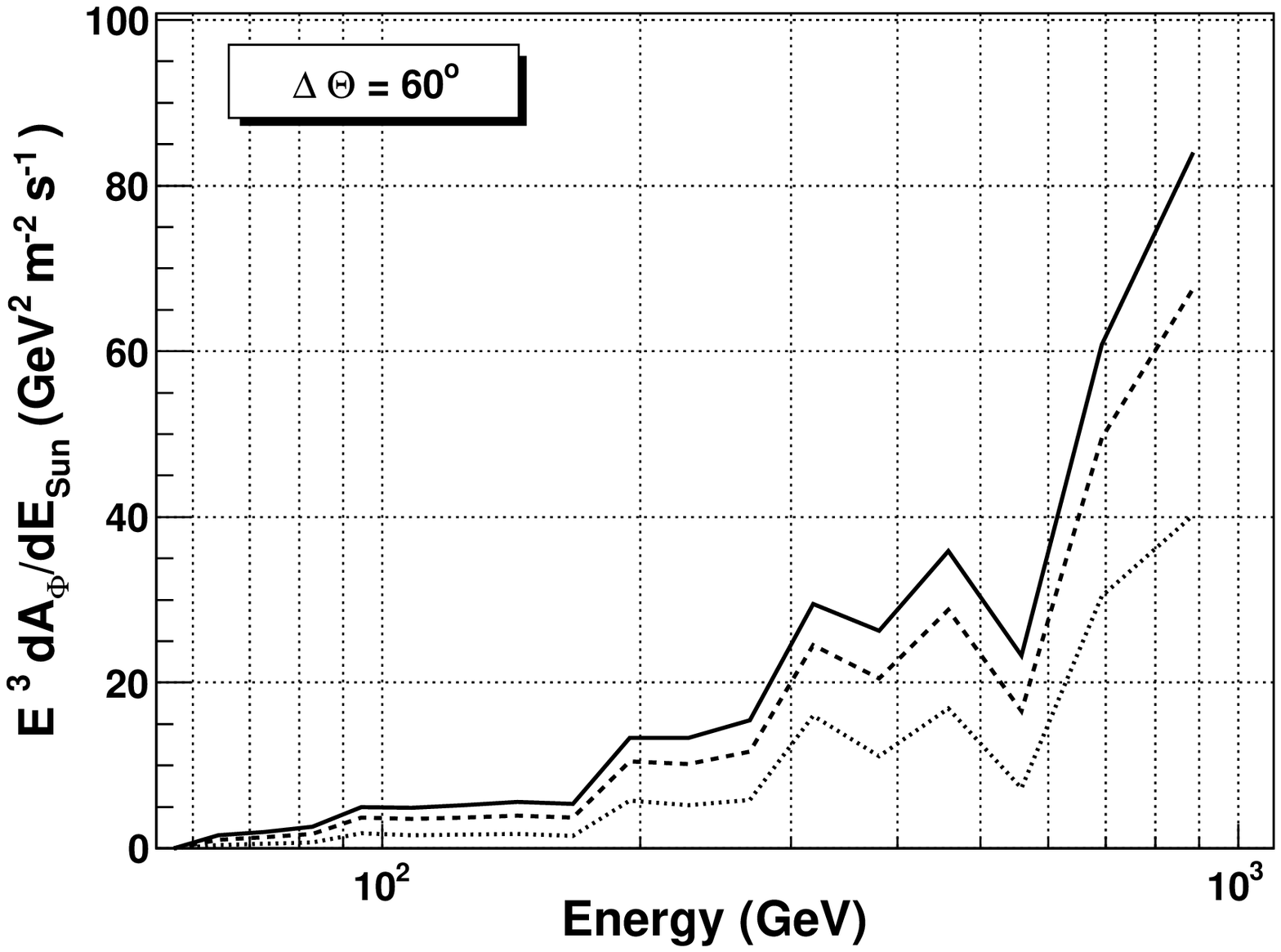}
\includegraphics[width=0.48\linewidth]{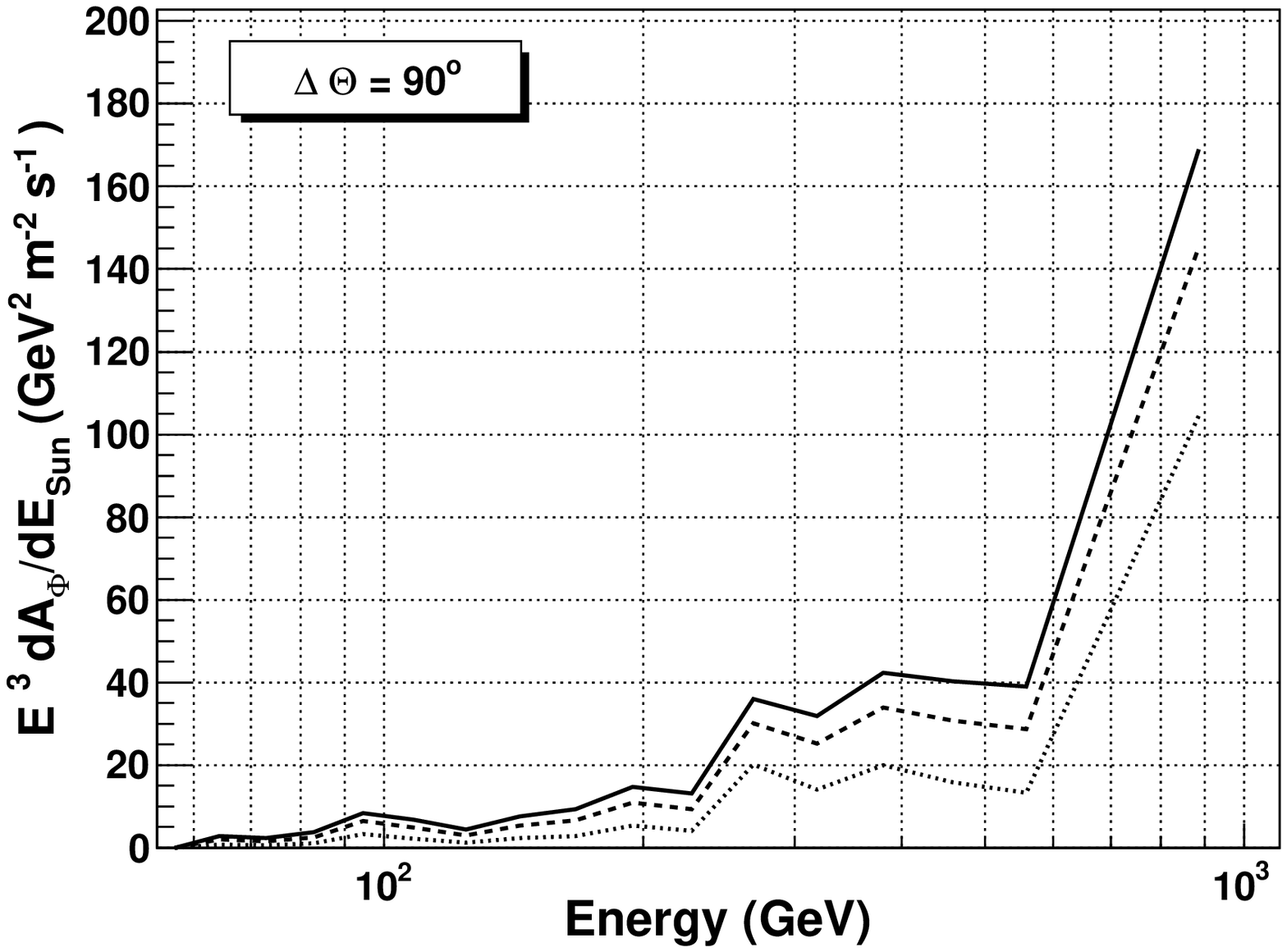}
\caption{Statistical upper limits at confidence levels of 
$68\%$ (dotted lines), $95\%$ (dashed lines) and $99\%$ 
(continuous lines) for the CRE flux asymmetry between real and fake
Sun, evaluated in cones with angular radii 
$\Delta \Theta = 30\degrees$ (top left panel), $45\degrees$ 
(top right panel), $60 \degrees$ (bottom left panel) 
and $90\degrees$ (bottom right panel).\label{fig:e3upperlimits_fake}}
\end{figure*}

The previous analysis did not provide any evidence of a CRE flux 
excess from the Sun with respect to the fake Sun, so we
set statistical upper limits on this signal by following the approach outlined in
Ref.~\cite{Cowan1998} (pp.~136-139).
In each energy bin, the measured flux asymmetry 
$dA_{\Phi}(E | \Delta \Theta)/dE$ can be seen as a realization 
of a Gaussian random variable. Assuming the hypothesis of a
CRE flux excess from the Sun, its expectation value must be
non-negative:

\begin{equation}  
\langle \frac{dA_{\Phi}(E | \Delta \Theta)}{dE} \rangle \geq 0 
\end{equation}
To set upper limits on $\langle dA_{\Phi}(E | \Delta \Theta)/dE \rangle$ 
we implemented the Bayesian method described in
Ref.~\cite{Cowan1998}, assuming a uniform prior density. The 
$\sigma$ of the Gaussian probability distribution function
associated with each measurement of $dA_{\Phi}(E | \Delta \Theta)/dE$
is the statistical error associated with the measurement.  

Fig.~\ref{fig:e3upperlimits_fake} shows the statistical 
upper limits on CRE flux asymmetries at different confidence 
levels, considering different cones centered on the Sun 
with angular radii ranging from $30\degrees$ to $90\degrees$.
In Table~\ref{tab:ul1} the values of the upper limits at
$95\%$ confidence level for the flux asymmetry 
$dA_{\Phi}(E | \Delta \Theta)/dE$ in the different ROIs
are summarized. The calculated
upper limits are also expressed in terms of fractions
of the CRE flux from the region of the fake Sun.

\begin{table*}
\begin{tabular}{||c||c|c||c|c||c|c||c|c||}
\hline
\hline
 & \multicolumn{2}{|c||}{$\Delta \Theta = 30\degrees$} 
 & \multicolumn{2}{|c||}{$\Delta \Theta = 45\degrees$} 
 & \multicolumn{2}{|c||}{$\Delta \Theta = 60\degrees$} 
 & \multicolumn{2}{|c||}{$\Delta \Theta = 90\degrees$} \\
\hline
Energy  & Flux UL & Fractional & Flux UL & Fractional & Flux UL & Fractional & Flux UL & Fractional \\
(GeV)   & ($\units{GeV^{-1}m^{-2}s^{-1}}$) & UL &
($\units{GeV^{-1}m^{-2}s^{-1}}$) & UL &
($\units{GeV^{-1}m^{-2}s^{-1}}$) & UL & 
($\units{GeV^{-1}m^{-2}s^{-1}}$) & UL \\	
\hline
\hline
$      60.4 -      68.2 $ & $ 3.508 \cdot 10^{-6} $ & $ 0.008 $ & $ 4.650 \cdot 10^{-6} $ & $ 0.005 $ & $ 3.934 \cdot 10^{-6} $ & $ 0.002 $ & $ 7.506 \cdot 10^{-6} $ & $ 0.002 $ \\
$      68.2 -      77.4 $ & $ 2.114 \cdot 10^{-6} $ & $ 0.007 $ & $ 2.496 \cdot 10^{-6} $ & $ 0.004 $ & $ 3.518 \cdot 10^{-6} $ & $ 0.003 $ & $ 4.096 \cdot 10^{-6} $ & $ 0.002 $ \\
$      77.4 -      88.1 $ & $ 2.744 \cdot 10^{-6} $ & $ 0.013 $ & $ 5.506 \cdot 10^{-6} $ & $ 0.012 $ & $ 3.131 \cdot 10^{-6} $ & $ 0.004 $ & $ 4.555 \cdot 10^{-6} $ & $ 0.003 $ \\
$      88.1 - 101 $ & $ 2.516 \cdot 10^{-6} $ & $ 0.019 $ & $ 5.127 \cdot 10^{-6} $ & $ 0.018 $ & $ 4.400 \cdot 10^{-6} $ & $ 0.009 $ & $ 7.696 \cdot 10^{-6} $ & $ 0.008 $ \\
$ 101 - 116 $ & $ 2.190 \cdot 10^{-6} $ & $ 0.024 $ & $ 1.963 \cdot 10^{-6} $ & $ 0.010 $ & $ 2.779 \cdot 10^{-6} $ & $ 0.008 $ & $ 3.845 \cdot 10^{-6} $ & $ 0.006 $ \\
$ 116 - 133 $ & $ 1.026 \cdot 10^{-6} $ & $ 0.017 $ & $ 9.091 \cdot 10^{-7} $ & $ 0.007 $ & $ 1.935 \cdot 10^{-6} $ & $ 0.009 $ & $ 1.583 \cdot 10^{-6} $ & $ 0.004 $ \\
$ 133 - 154 $ & $ 1.471 \cdot 10^{-6} $ & $ 0.039 $ & $ 1.261 \cdot 10^{-6} $ & $ 0.015 $ & $ 1.337 \cdot 10^{-6} $ & $ 0.010 $ & $ 1.795 \cdot 10^{-6} $ & $ 0.006 $ \\
$ 154 - 180 $ & $ 5.671 \cdot 10^{-7} $ & $ 0.021 $ & $ 5.297 \cdot 10^{-7} $ & $ 0.009 $ & $ 7.971 \cdot 10^{-7} $ & $ 0.008 $ & $ 1.435 \cdot 10^{-6} $ & $ 0.007 $ \\
$ 180 - 210 $ & $ 8.580 \cdot 10^{-7} $ & $ 0.054 $ & $ 1.348 \cdot 10^{-6} $ & $ 0.039 $ & $ 1.419 \cdot 10^{-6} $ & $ 0.025 $ & $ 1.489 \cdot 10^{-6} $ & $ 0.013 $ \\
$ 210 - 246 $ & $ 1.252 \cdot 10^{-6} $ & $ 0.133 $ & $ 9.240 \cdot 10^{-7} $ & $ 0.045 $ & $ 8.574 \cdot 10^{-7} $ & $ 0.024 $ & $ 7.953 \cdot 10^{-7} $ & $ 0.011 $ \\
$ 246 - 291 $ & $ 2.905 \cdot 10^{-7} $ & $ 0.049 $ & $ 4.411 \cdot 10^{-7} $ & $ 0.034 $ & $ 6.033 \cdot 10^{-7} $ & $ 0.027 $ & $ 1.556 \cdot 10^{-6} $ & $ 0.036 $ \\
$ 291 - 346 $ & $ 3.946 \cdot 10^{-7} $ & $ 0.111 $ & $ 5.473 \cdot 10^{-7} $ & $ 0.073 $ & $ 7.581 \cdot 10^{-7} $ & $ 0.059 $ & $ 7.778 \cdot 10^{-7} $ & $ 0.030 $ \\
$ 346 - 415 $ & $ 2.457 \cdot 10^{-7} $ & $ 0.115 $ & $ 2.925 \cdot 10^{-7} $ & $ 0.064 $ & $ 3.715 \cdot 10^{-7} $ & $ 0.048 $ & $ 6.160 \cdot 10^{-7} $ & $ 0.040 $ \\
$ 415 - 503 $ & $ 1.567 \cdot 10^{-7} $ & $ 0.140 $ & $ 2.057 \cdot 10^{-7} $ & $ 0.085 $ & $ 2.979 \cdot 10^{-7} $ & $ 0.072 $ & $ 3.187 \cdot 10^{-7} $ & $ 0.038 $ \\
$ 503 - 615 $ & $ 6.318 \cdot 10^{-8} $ & $ 0.094 $ & $ 7.052 \cdot 10^{-8} $ & $ 0.049 $ & $ 9.512 \cdot 10^{-8} $ & $ 0.041 $ & $ 1.644 \cdot 10^{-7} $ & $ 0.037 $ \\
$ 615 - 772 $ & $ 7.297 \cdot 10^{-8} $ & $ 0.251 $ & $ 1.047 \cdot 10^{-7} $ & $ 0.169 $ & $ 1.486 \cdot 10^{-7} $ & $ 0.134 $ & $ 2.501 \cdot 10^{-7} $ & $ 0.111 $ \\
$ 772 - 1000 $ & $ 6.097 \cdot 10^{-8} $ & $ 0.493 $ & $ 5.951 \cdot
10^{-8} $ & $ 0.192 $ & $ 9.720 \cdot 10^{-8} $ & $ 0.178 $ & $ 2.088
\cdot 10^{-7} $ & $ 0.196 $ \\
\hline
\hline
\end{tabular}
\caption{Statistical upper limits at $95\%$ confidence level on the
CRE flux asymmetries between the real and the fake Sun, evaluated
in cones of angular radii $\Delta \Theta = 30\degrees, 45\degrees,
60\degrees, 90\degrees$. The upper limits are also
expressed in terms of fractions of the CRE flux from the fake Sun.}
\label{tab:ul1}
\end{table*}

\subsection{Comparison with an isotropic flux}
\label{sec:isotropic}

The second approach used in this analysis is based on the 
event-shuffling technique
employed in Ref.~\cite{Ackermann:2010ip}, which was used
to build a simulated sample of isotropic CREs starting 
from the real events. Simulated events are built
by randomly coupling the arrival times and the arrival 
directions (in local instrument coordinates) of real events. 
The simulated event sample used for this analysis
is the same used in Ref.~\cite{Ackermann:2010ip} and 
is $100$ times larger than the real one.

In this case, given the angular radius $\Delta \Theta$ of a 
cone centered on the Sun, we evaluated the count 
differences between real and simulated CREs as:

\begin{equation}
\Delta N (E|\Delta \Theta) = 
N_{real}(E | \Delta \Theta) - \alpha (E) N_{sim}(E | \Delta \Theta)
\label{eq:counts}
\end{equation}
where $N_{real}(E|\Delta \Theta)$ and $N_{sim}(E|\Delta \Theta)$
are respectively the number of CRE events in the real 
and simulated data sets with energy $E$ and arrival directions 
within the selected cone. The parameter $\alpha(E)$ in 
Eq.~\ref{eq:counts} is a normalization factor. 
If $N_{real}(E)$ and $N_{sim}(E)$ are the total numbers of real 
and simulated events with energy $E$, 
then $\alpha(E)=N_{real}(E)/N_{sim}(E)$~\footnote{The factor $\alpha(E)$ is not exactly equal to
$1/100$ (that is the exact ratio between the sizes of the
real and simulated event samples), because in the 
randomization process only the overall number of simulated 
events is fixed, but the simulated events in individual 
energy bins can change. However, since the size of the simulated 
data sample is large, $\alpha(E) \simeq 1/100$ in each energy bin.}. 

The count difference is finally converted into a flux asymmetry
according to the following equation:

\begin{equation}
\cfrac{dA'_{\Phi}(E | \Delta \Theta)}{dE} =  \cfrac{1}{\Delta E} 
\cdot   
\cfrac{ \Delta N(E|\Delta \Theta)  (1-c(E))}
{\mathcal{E} (E | \Delta \Theta)} 
\label{eq:asymmetry2}
\end{equation}
where $c(E)$ is the residual contamination and
$\mathcal{E}(E|\Delta\Theta)$ is the exposure of the 
selected sky region, which is evaluated from 
the instrument effective area and the live times
of the pixels belonging to that sky region.

We emphasize that the two approaches used in this work are
complementary, but not fully equivalent. In the
first case the variable $dA_{\Phi}(E | \Delta \Theta)/dE$ is built using
real events from different directions (real Sun and fake Sun).
On the other hand, in the second case, the variable 
$dA'_{\Phi}(E | \Delta \Theta)/dE$ is built using real and 
simulated events from the same region of the sky. In both 
cases the goal of the analysis is to compare the 
CRE flux from the Sun with the average CRE flux.
In the first case, the reference flux is evaluated looking 
at real events from the fake Sun region, while in the 
second case it is evaluated simulating an isotropic flux 
and looking at simulated events from the Sun region.

The second approach, however, excludes potential  
systematic biases when calculating flux differences. In particular, 
to evaluate the flux from a given region of the sky requires 
knowledge of the exposure, which in turn depends
on the effective area of the detector and on the observation
live time. The effective area is calculated from Monte Carlo
simulations and thus could be affected by systematics such as 
variations correlated with time and spacecraft position 
or miscalculations of its dependence on instrument coordinates. 
When evaluating the flux asymmetry $dA_{\Phi}(E | \Delta \Theta)/dE$ according 
to Eq.~\ref{eq:asymmetry1}, the systematic uncertainties involved in the 
evaluation of the two terms could be different,
and the result could be biased. On the other hand, when evaluating
the flux difference $dA'_{\Phi}(E | \Delta \Theta)/dE$ from Eq.~\ref{eq:asymmetry2},
inaccuracies in the effective area calculation can only
result in a scale error on the flux difference.

\begin{figure*}
\includegraphics[width=0.48\linewidth]{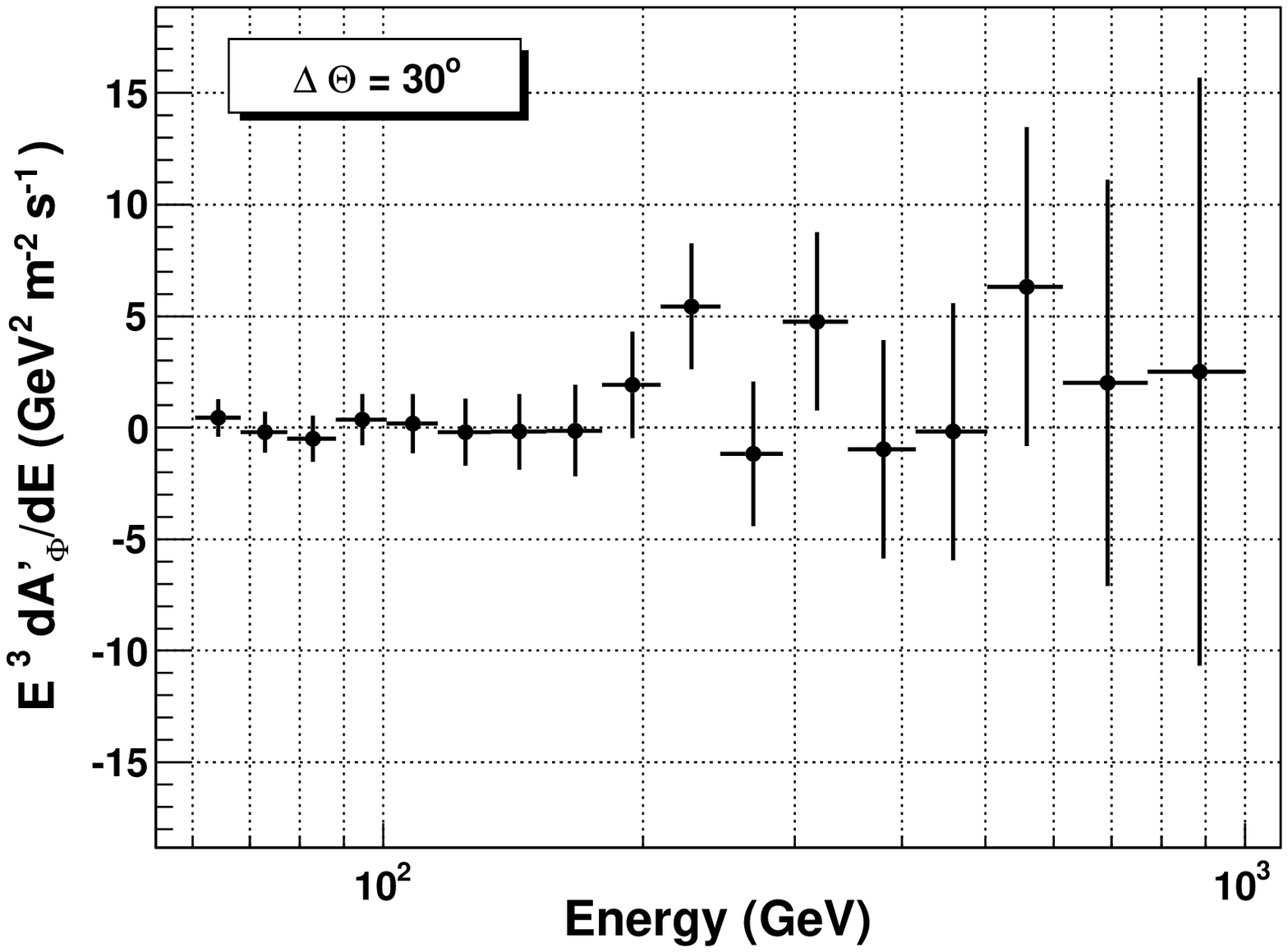}
\includegraphics[width=0.48\linewidth]{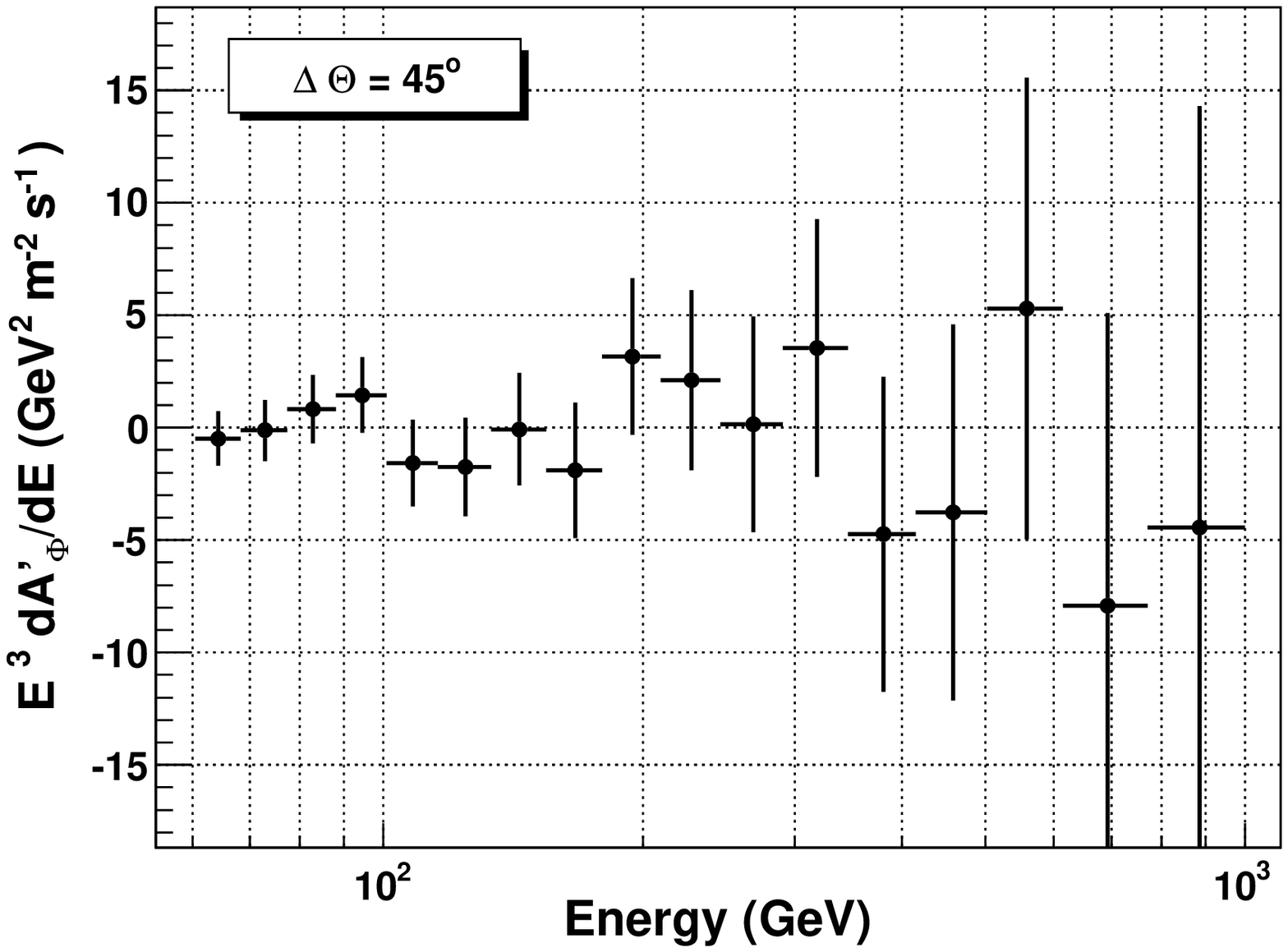}
\includegraphics[width=0.48\linewidth]{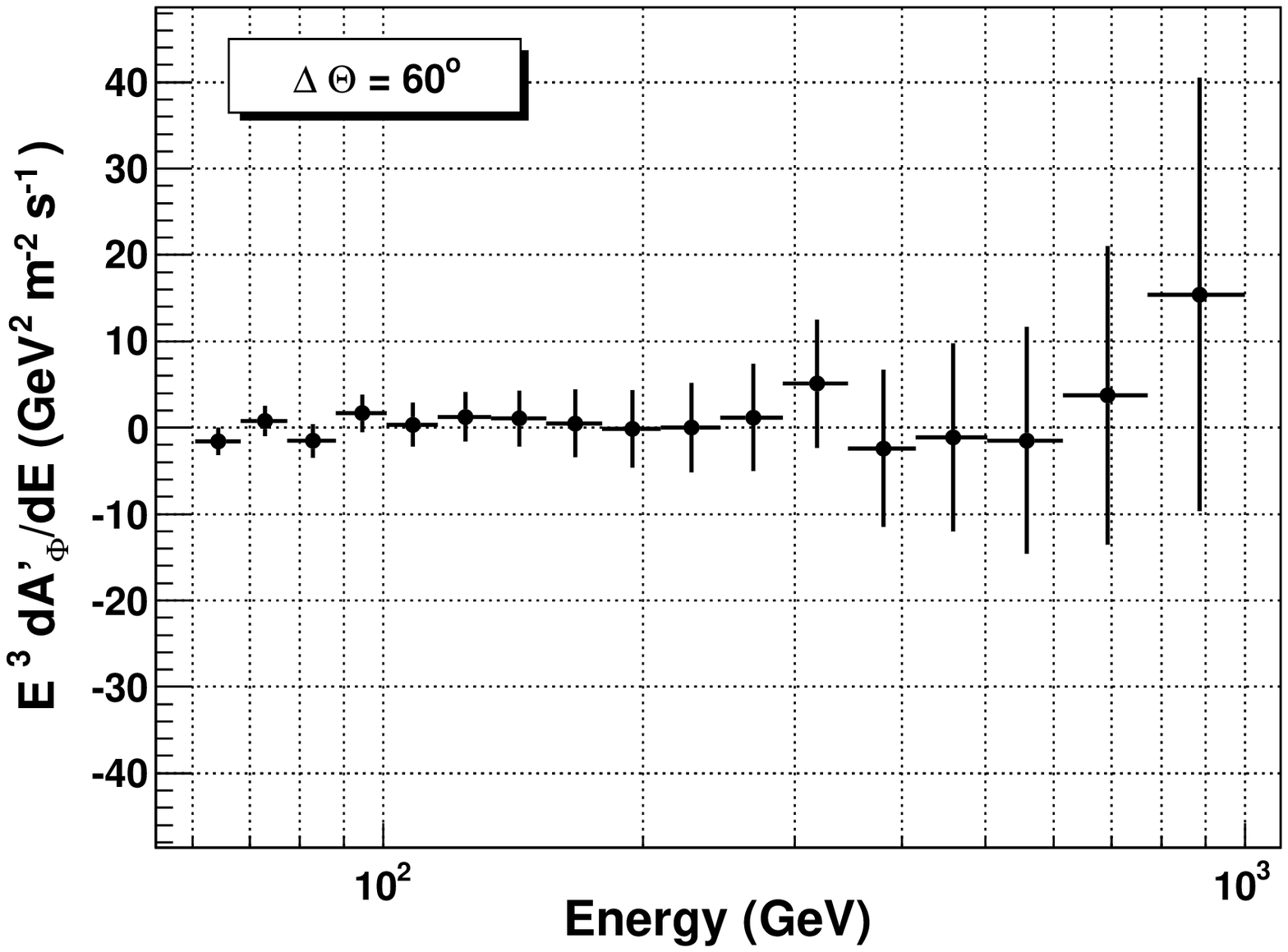}
\includegraphics[width=0.48\linewidth]{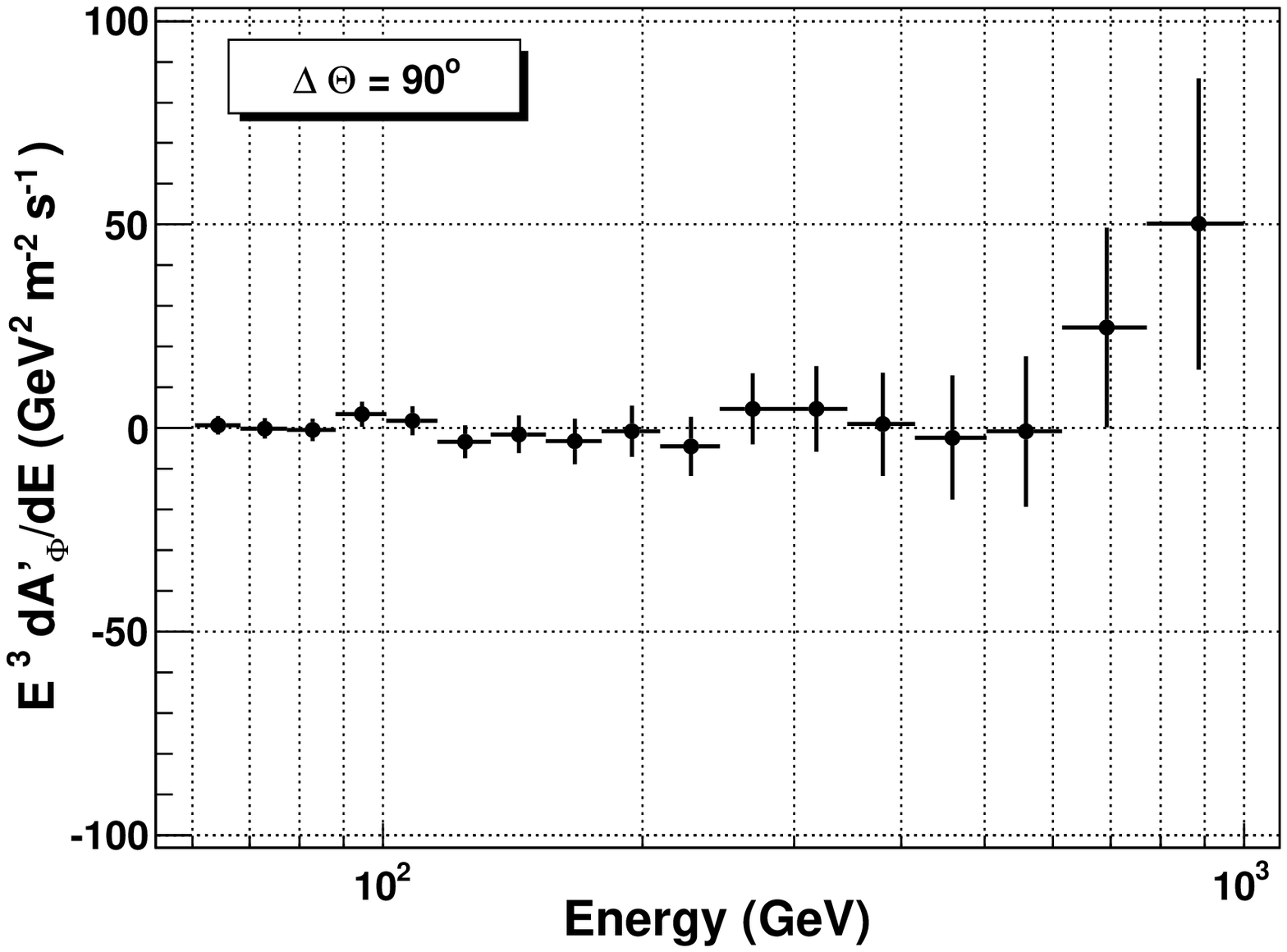}
\caption{Differential flux asymmetry between real and simulated events
from the Sun evaluated in cones with angular radii 
$\Delta \Theta = 30\degrees$ (top left panel), $45\degrees$ 
(top right panel), $60 \degrees$ (bottom left panel) 
and $90\degrees$ (bottom right panel).  
Only statistical error bars are shown.
\label{fig:e3counts}}
\end{figure*}

To determine whether the real counts differ significantly
from the simulated ones, we performed a hypothesis test 
following the prescriptions of Ref.~\cite{Li:1983fv}. 
Denoting with $N_{real}(E | \Delta \Theta)$ and 
$N_{sim}(E | \Delta\Theta)$ the real and the simulated 
counts in the energy bin $E$, the null hypothesis for 
this analysis is that 
$N_{real}(E | \Delta \Theta) = \alpha(E) N_{sim}(E | \Delta \Theta)$.
Following Ref.~\cite{Li:1983fv}, we evaluated 
the significance in each energy bin as:

\begin{equation}
\begin{split}
& S(E | \Delta \Theta) = \pm \sqrt{2} \left\{  
N_{real}(E | \Delta \Theta)  
\ln \left[ \cfrac{1 + \alpha(E)}{\alpha(E)} \right. \right. \\& 
\left. \cfrac{N_{real}(E | \Delta \Theta)}  
{N_{real}(E | \Delta \Theta) + N_{sim}(E | \Delta \Theta)} 
 \right] + 
N_{sim}(E | \Delta \Theta) \\& 
\left. \ln \left[ \left(1 + \alpha(E) \right)
\cfrac{N_{sim}(E | \Delta \Theta)}
{N_{real}(E | \Delta \Theta) + N_{sim}(E | \Delta \Theta)}
\right] 
\right\}^{1/2}
\end{split}
\label{eq:significance}
\end{equation}
with the convention of choosing the $+$ sign if 
$N_{real} (E | \Delta \Theta ) > \alpha(E) N_{sim} (E | \Delta \Theta)$
and the $-$ sign if
$N_{real} (E | \Delta \Theta ) < \alpha(E) N_{sim} (E | \Delta \Theta)$.

The significance values evaluated from Eq.~\ref{eq:significance}
can be converted into probability values. In particular, since 
$S^{2}$ is a random variable following a $\chi^{2}$ distribution
with $1$ degree of freedom~\cite{Li:1983fv}, one can easily evaluate
the probability of observing a value of $S^{2}$ larger than
the one observed. 
In Table~\ref{tab:prob2} the maximum deviations from the null
flux asymmetries are shown for each ROI used for our analysis,
together with the corresponding probabilities of finding
larger deviations. The last column of the table shows
the probabilities of finding, in each ROI, 
at least one energy bin with a flux asymmetry larger than
the maximum observed value. Again, these probabilities
were calculated assuming that the flux asymmetries
measured in each of the $17$ energy bins used for our 
analysis are uncorrelated. The observed deviations
from the null flux asymmetries are statistically insignificant.

\begin{table*}[!ht]
\begin{tabular}{||c||c|c|c||}
\hline
Angular radius & Maximum deviation ($S_{max}$) & $P(|S_{max}|)$ & 
$P(|S|>|S_{max}|)$ \\
\hline
$30 \degrees$ &  1.925 & 0.054 & 0.611 \\
\hline
$45 \degrees$ &  0.419 & 0.675 & 1.000 \\
\hline
$60 \degrees$ & -0.654 & 0.513 & 1.000 \\
\hline
$90 \degrees$ &  1.026 & 0.305 & 0.998 \\
\hline
\end{tabular}
\caption{For each cone used for the flux asymmetry analysis 
the maximum deviation (either positive or negative) in terms of
significance from the null value are shown with the corresponding 
probability of finding a larger significance value. 
The last column shows the probability of finding at least one energy 
bin a with larger flux asymmetry than the maximum observed value. 
\label{tab:prob2}}
\end{table*}

The count differences were converted into
flux differences according to Eq.~\ref{eq:asymmetry2}.
In Fig.~\ref{fig:e3counts} the asymmetry variable 
$dA'_{\Phi}(E|\Delta \Theta)/dE$ between the real and 
simulated fluxes from the Sun is shown for the four 
cones with angular radii of $30\degrees$, $45\degrees$, 
$60 \degrees$ and $90\degrees$. Again, no evidence of a
CRE signal from the Sun is observed. Similar results
are obtained when integral fluxes are analyzed.

\subsubsection{Evaluation of statistical upper limits 
on the CRE flux from the Sun}

\begin{figure*}
\includegraphics[width=0.48\linewidth]{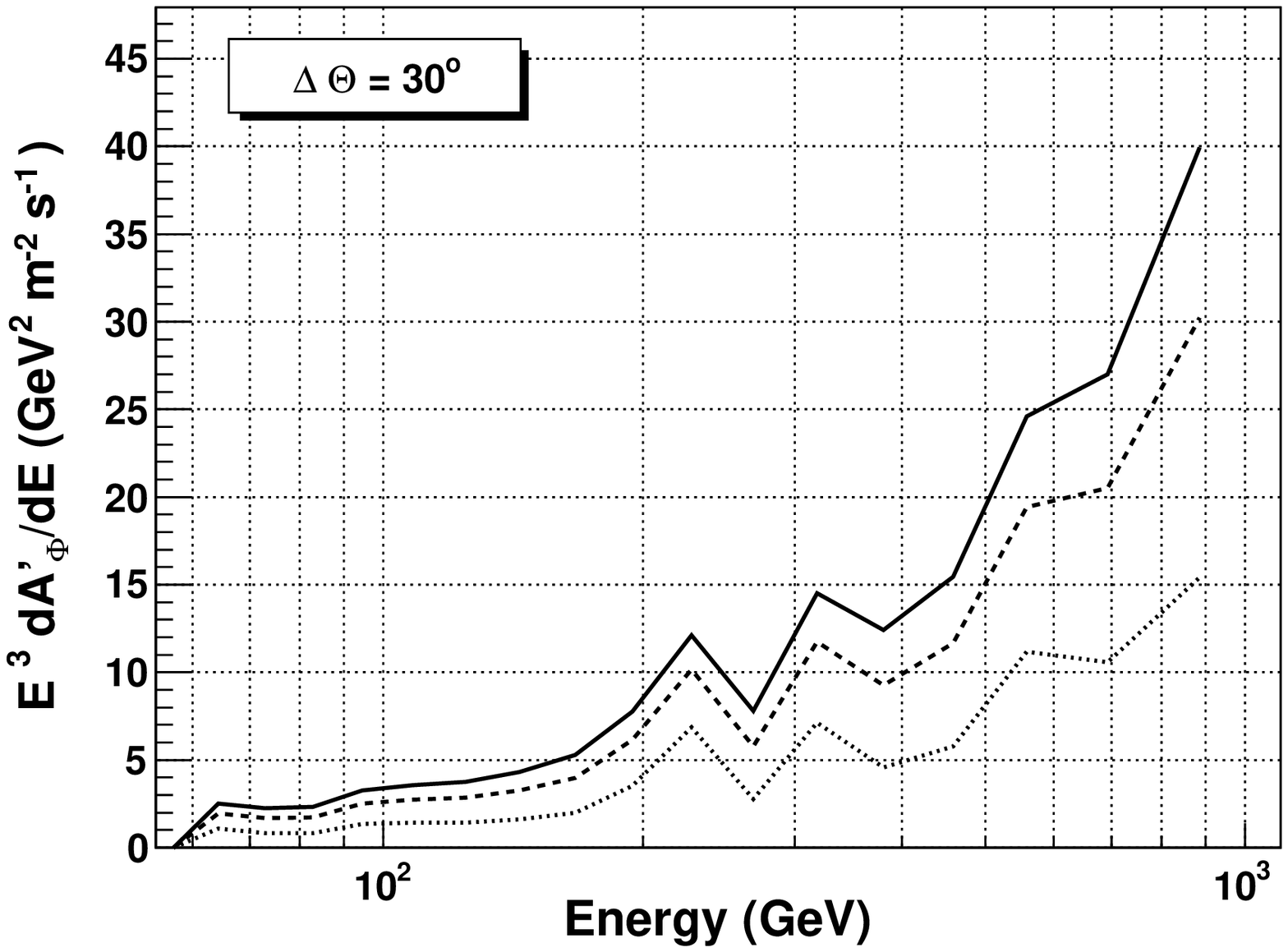}
\includegraphics[width=0.48\linewidth]{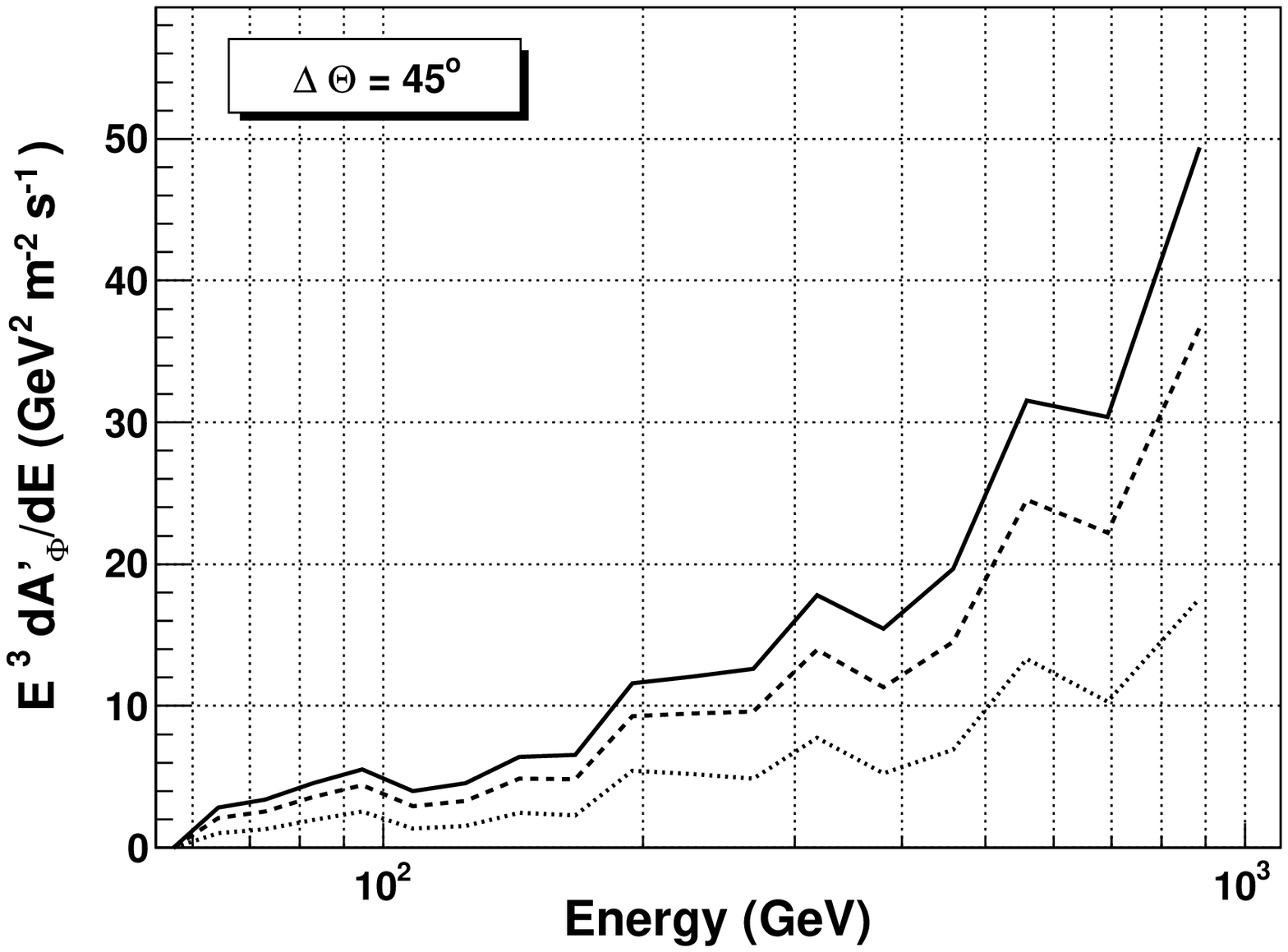}
\includegraphics[width=0.48\linewidth]{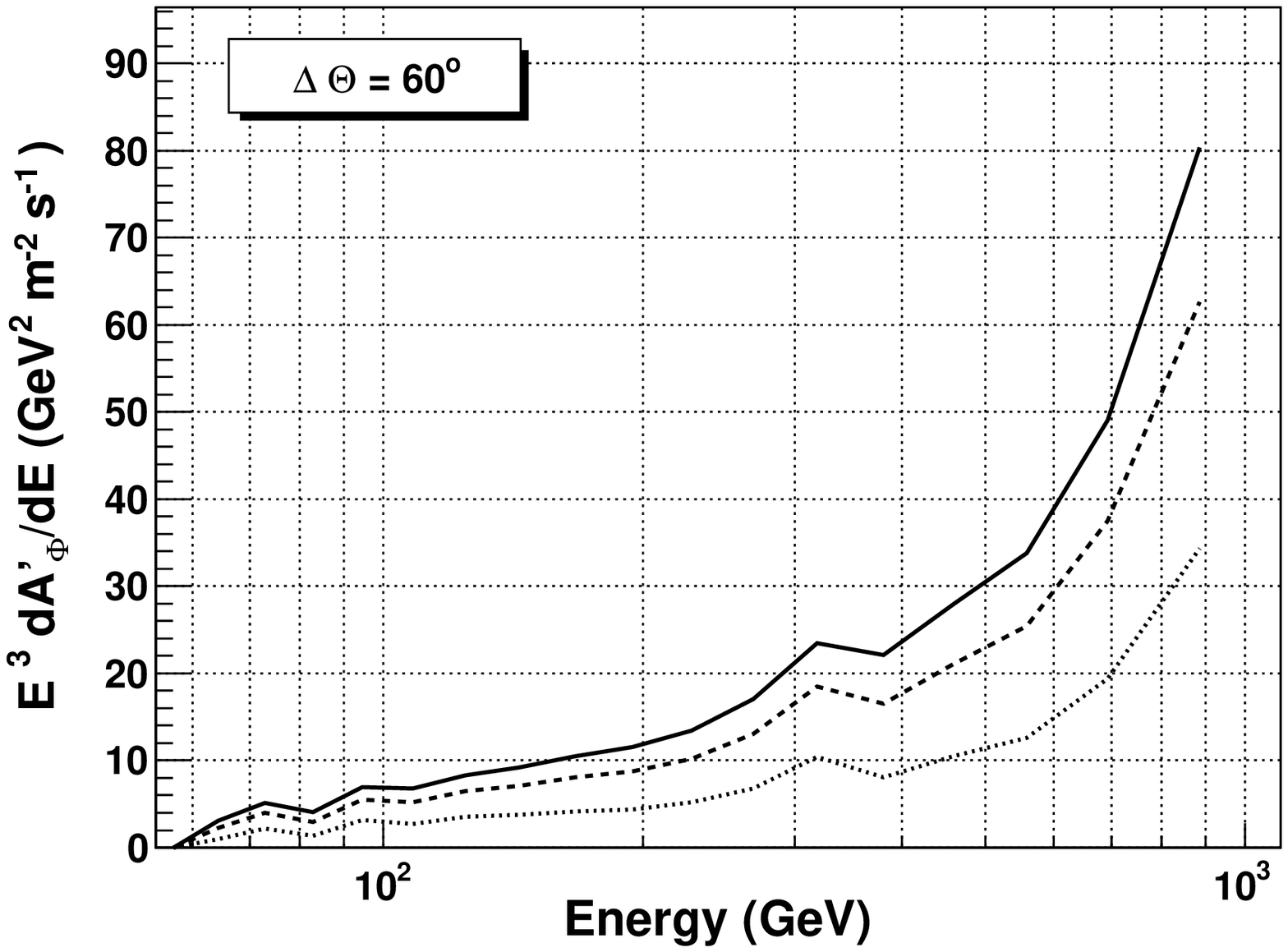}
\includegraphics[width=0.48\linewidth]{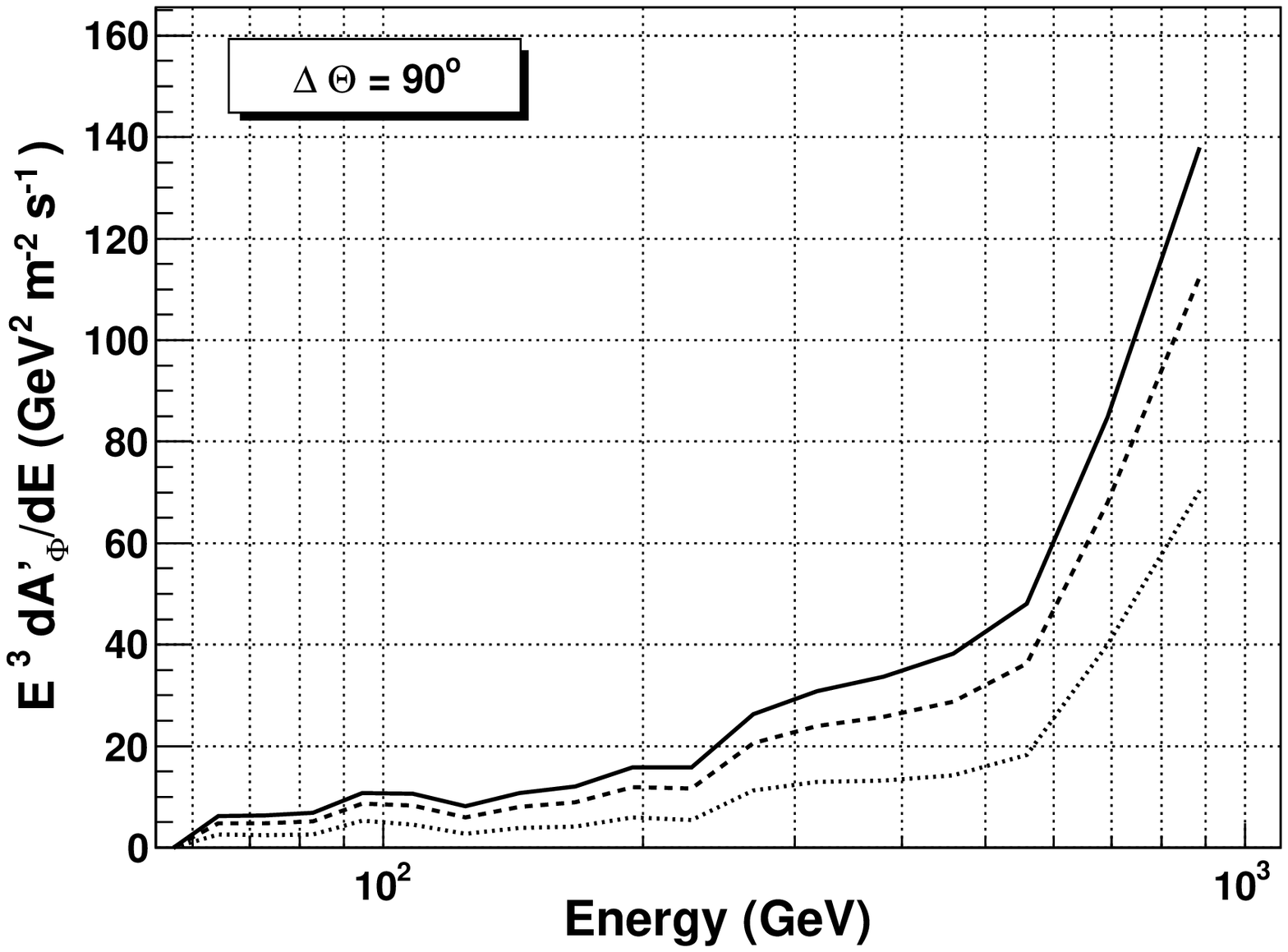}
\caption{Statistical upper limits at confidence levels of 
of $68\%$ (dotted lines), $95\%$ (dashed lines) and $99\%$ 
(continuous lines) for the CRE fluxes from the Sun, evaluated 
in cones with angular radii 
$\Delta \Theta = 30\degrees$ (top left panel), $45\degrees$ 
(top right panel), $60 \degrees$ (bottom left panel) 
and $90\degrees$ (bottom right panel).\label{fig:e3upperlimits}}
\end{figure*}

As in \S\ref{sec:fluxasym}, our analysis in this case does not find 
evidence of a CRE signal from the Sun, and so we set statistical 
upper limits on solar CRE fluxes.
Given a cone with angular radius $\Delta \Theta$ centered on the
Sun, the observed counts $N_{real}(E | \Delta \Theta)$ in the energy
bin $E$ can be seen as a realization of a Poisson random variable. 
Assuming the hypothesis of a CRE signal from the Sun, 
these counts will be the sum of a signal contribution, 
$N_{Sun}(E | \Delta \Theta)$, plus a background contribution,
$N_{bkg}(E | \Delta \Theta)$:

\begin{equation}
N_{real}(E | \Delta \Theta) = N_{Sun}(E | \Delta \Theta) + N_{bkg}(E | \Delta \Theta).
\end{equation}
Both $N_{Sun}(E | \Delta \Theta)$ and $N_{bkg}(E | \Delta \Theta)$ can
be seen as Poisson random variables. The only information available
about $N_{Sun}(E | \Delta \Theta)$ is that, in the hypothesis of a
CRE signal from the Sun, its average value (which is unknown) 
must be non-negative:

\begin{equation}
\langle N_{Sun}(E | \Delta \Theta) \rangle \geq 0.
\end{equation}
On the other hand, the average value of $N_{bkg}(E | \Delta \Theta)$ 
can be estimated from the randomized data sets, and is given by:

\begin{equation}
\langle N_{bkg}(E | \Delta \Theta) \rangle = \alpha(E) N_{sim} (E | \Delta \Theta) 
\end{equation}
where $\alpha(E)$ is the ratio between the total real and simulated
events in the energy bin $E$. 

The goal of our analysis is to evaluate an upper limit on the
average value $\langle N_{Sun}(E | \Delta \Theta) \rangle$, that will
be converted into an upper limit on the CRE flux from the Sun
after properly taking into account the detector acceptance and
the live time. For our calculation we implemented 
the Bayesian method with the assumption of a uniform prior density. 
The mathematical details of the method can be found 
in Ref.~\cite{Cowan1998} (pp.~139-142).

Fig.~\ref{fig:e3upperlimits} shows the statistical upper limits 
at different confidence levels on the CRE fluxes 
from different cones centered on the Sun and with 
angular radii ranging from $30\degrees$ to $90\degrees$.
The results are consistent with those shown in
Fig.~\ref{fig:e3upperlimits_fake}.
In Table~\ref{tab:ul2} the values of the upper limits at
$95\%$ confidence level for the flux asymmetry 
$dA'_{\Phi}(E | \Delta \Theta)/dE$ in the different ROIs
are summarized. The calculated
upper limits are also expressed in terms of fractional
excess with respect to the expected isotropic flux.

\begin{table*}
\begin{tabular}{||c||c|c||c|c||c|c||c|c||}
\hline
\hline
 & \multicolumn{2}{|c||}{$\Delta \Theta = 30\degrees$} 
 & \multicolumn{2}{|c||}{$\Delta \Theta = 45\degrees$} 
 & \multicolumn{2}{|c||}{$\Delta \Theta = 60\degrees$} 
 & \multicolumn{2}{|c||}{$\Delta \Theta = 90\degrees$} \\
\hline
Energy  & Flux UL & Fractional & Flux UL & Fractional & Flux UL & Fractional & Flux UL & Fractional \\
(GeV)   & ($\units{GeV^{-1}m^{-2}s^{-1}}$) & UL &
($\units{GeV^{-1}m^{-2}s^{-1}}$) & UL &
($\units{GeV^{-1}m^{-2}s^{-1}}$) & UL & 
($\units{GeV^{-1}m^{-2}s^{-1}}$) & UL \\	
\hline
\hline
$      60.4 -      68.2 $ & $ 7.344 \cdot 10^{-6} $ & $ 0.017 $ & $ 7.865 \cdot 10^{-6} $ & $ 0.008 $ & $ 8.390 \cdot 10^{-6} $ & $ 0.005 $ & $ 1.823 \cdot 10^{-5} $ & $ 0.006 $ \\
$      68.2 -      77.4 $ & $ 4.366 \cdot 10^{-6} $ & $ 0.015 $ & $ 6.652 \cdot 10^{-6} $ & $ 0.010 $ & $ 1.033 \cdot 10^{-5} $ & $ 0.009 $ & $ 1.245 \cdot 10^{-5} $ & $ 0.006 $ \\
$      77.4 -      88.1 $ & $ 3.060 \cdot 10^{-6} $ & $ 0.015 $ & $ 6.279 \cdot 10^{-6} $ & $ 0.014 $ & $ 5.237 \cdot 10^{-6} $ & $ 0.007 $ & $ 9.069 \cdot 10^{-6} $ & $ 0.006 $ \\
$      88.1 - 101 $ & $ 2.987 \cdot 10^{-6} $ & $ 0.022 $ & $ 5.211 \cdot 10^{-6} $ & $ 0.018 $ & $ 6.548 \cdot 10^{-6} $ & $ 0.013 $ & $ 1.033 \cdot 10^{-5} $ & $ 0.010 $ \\
$ 101 - 116 $ & $ 2.162 \cdot 10^{-6} $ & $ 0.023 $ & $ 2.296 \cdot 10^{-6} $ & $ 0.012 $ & $ 4.093 \cdot 10^{-6} $ & $ 0.012 $ & $ 6.541 \cdot 10^{-6} $ & $ 0.010 $ \\
$ 116 - 133 $ & $ 1.471 \cdot 10^{-6} $ & $ 0.025 $ & $ 1.710 \cdot 10^{-6} $ & $ 0.013 $ & $ 3.363 \cdot 10^{-6} $ & $ 0.015 $ & $ 3.091 \cdot 10^{-6} $ & $ 0.007 $ \\
$ 133 - 154 $ & $ 1.090 \cdot 10^{-6} $ & $ 0.029 $ & $ 1.631 \cdot 10^{-6} $ & $ 0.020 $ & $ 2.384 \cdot 10^{-6} $ & $ 0.017 $ & $ 2.683 \cdot 10^{-6} $ & $ 0.010 $ \\
$ 154 - 180 $ & $ 8.568 \cdot 10^{-7} $ & $ 0.033 $ & $ 1.034 \cdot 10^{-6} $ & $ 0.019 $ & $ 1.729 \cdot 10^{-6} $ & $ 0.018 $ & $ 1.912 \cdot 10^{-6} $ & $ 0.010 $ \\
$ 180 - 210 $ & $ 8.360 \cdot 10^{-7} $ & $ 0.052 $ & $ 1.257 \cdot 10^{-6} $ & $ 0.037 $ & $ 1.185 \cdot 10^{-6} $ & $ 0.020 $ & $ 1.614 \cdot 10^{-6} $ & $ 0.014 $ \\
$ 210 - 246 $ & $ 8.587 \cdot 10^{-7} $ & $ 0.087 $ & $ 7.968 \cdot 10^{-7} $ & $ 0.038 $ & $ 8.626 \cdot 10^{-7} $ & $ 0.024 $ & $ 9.827 \cdot 10^{-7} $ & $ 0.014 $ \\
$ 246 - 291 $ & $ 2.983 \cdot 10^{-7} $ & $ 0.050 $ & $ 4.946 \cdot 10^{-7} $ & $ 0.038 $ & $ 6.749 \cdot 10^{-7} $ & $ 0.031 $ & $ 1.062 \cdot 10^{-6} $ & $ 0.024 $ \\
$ 291 - 346 $ & $ 3.630 \cdot 10^{-7} $ & $ 0.100 $ & $ 4.318 \cdot 10^{-7} $ & $ 0.056 $ & $ 5.716 \cdot 10^{-7} $ & $ 0.043 $ & $ 7.422 \cdot 10^{-7} $ & $ 0.028 $ \\
$ 346 - 415 $ & $ 1.681 \cdot 10^{-7} $ & $ 0.074 $ & $ 2.048 \cdot 10^{-7} $ & $ 0.043 $ & $ 2.997 \cdot 10^{-7} $ & $ 0.038 $ & $ 4.661 \cdot 10^{-7} $ & $ 0.030 $ \\
$ 415 - 503 $ & $ 1.205 \cdot 10^{-7} $ & $ 0.103 $ & $ 1.504 \cdot 10^{-7} $ & $ 0.060 $ & $ 2.176 \cdot 10^{-7} $ & $ 0.051 $ & $ 2.979 \cdot 10^{-7} $ & $ 0.035 $ \\
$ 503 - 615 $ & $ 1.113 \cdot 10^{-7} $ & $ 0.184 $ & $ 1.409 \cdot 10^{-7} $ & $ 0.108 $ & $ 1.457 \cdot 10^{-7} $ & $ 0.066 $ & $ 2.082 \cdot 10^{-7} $ & $ 0.047 $ \\
$ 615 - 772 $ & $ 6.154 \cdot 10^{-8} $ & $ 0.197 $ & $ 6.662 \cdot 10^{-8} $ & $ 0.098 $ & $ 1.125 \cdot 10^{-7} $ & $ 0.097 $ & $ 2.038 \cdot 10^{-7} $ & $ 0.088 $ \\
$ 772 - 1000 $ & $ 4.344 \cdot 10^{-8} $ & $ 0.286 $ & $ 5.276 \cdot 10^{-8} $ & $ 0.162 $ & $ 9.005 \cdot 10^{-8} $ & $ 0.161 $ & $ 1.617 \cdot 10^{-7} $ & $ 0.145 $ \\
\hline
\hline
\end{tabular}
\caption{Statistical upper limits at $95\%$ confidence level on the
CRE flux asymmetries between the real and the simulated Sun, 
generated by the event shuffling technique, evaluated
in cones of angular radii $\Delta \Theta = 30\degrees, 45\degrees,
60\degrees, 90\degrees$. The upper limits are also
expressed in terms of fractions of the isotropic CRE flux 
from the simulated Sun.}
\label{tab:ul2}
\end{table*}

\subsubsection{Spherical harmonics analysis}

\begin{figure*}
\includegraphics[width=0.48\linewidth]{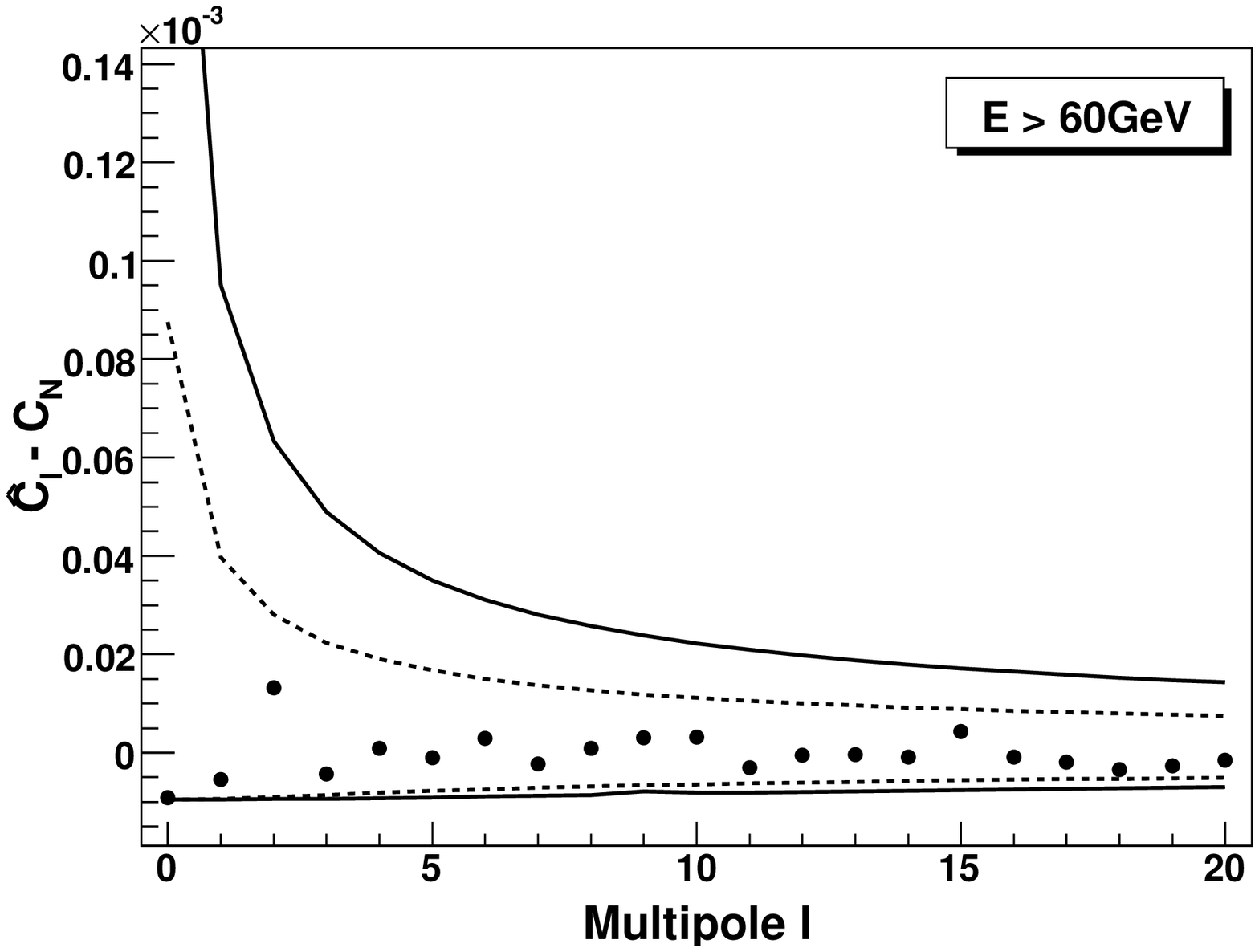}
\includegraphics[width=0.48\linewidth]{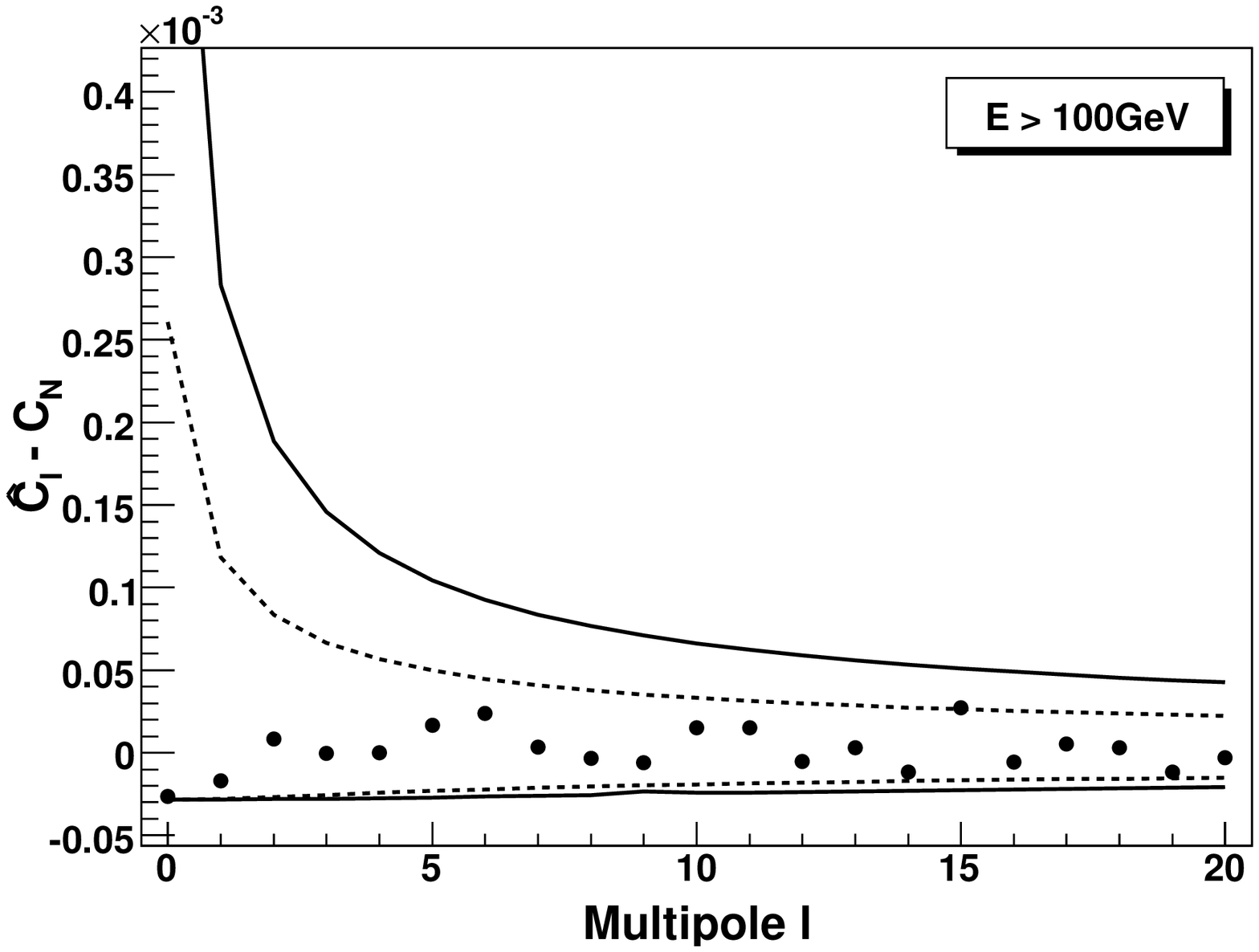}
\includegraphics[width=0.48\linewidth]{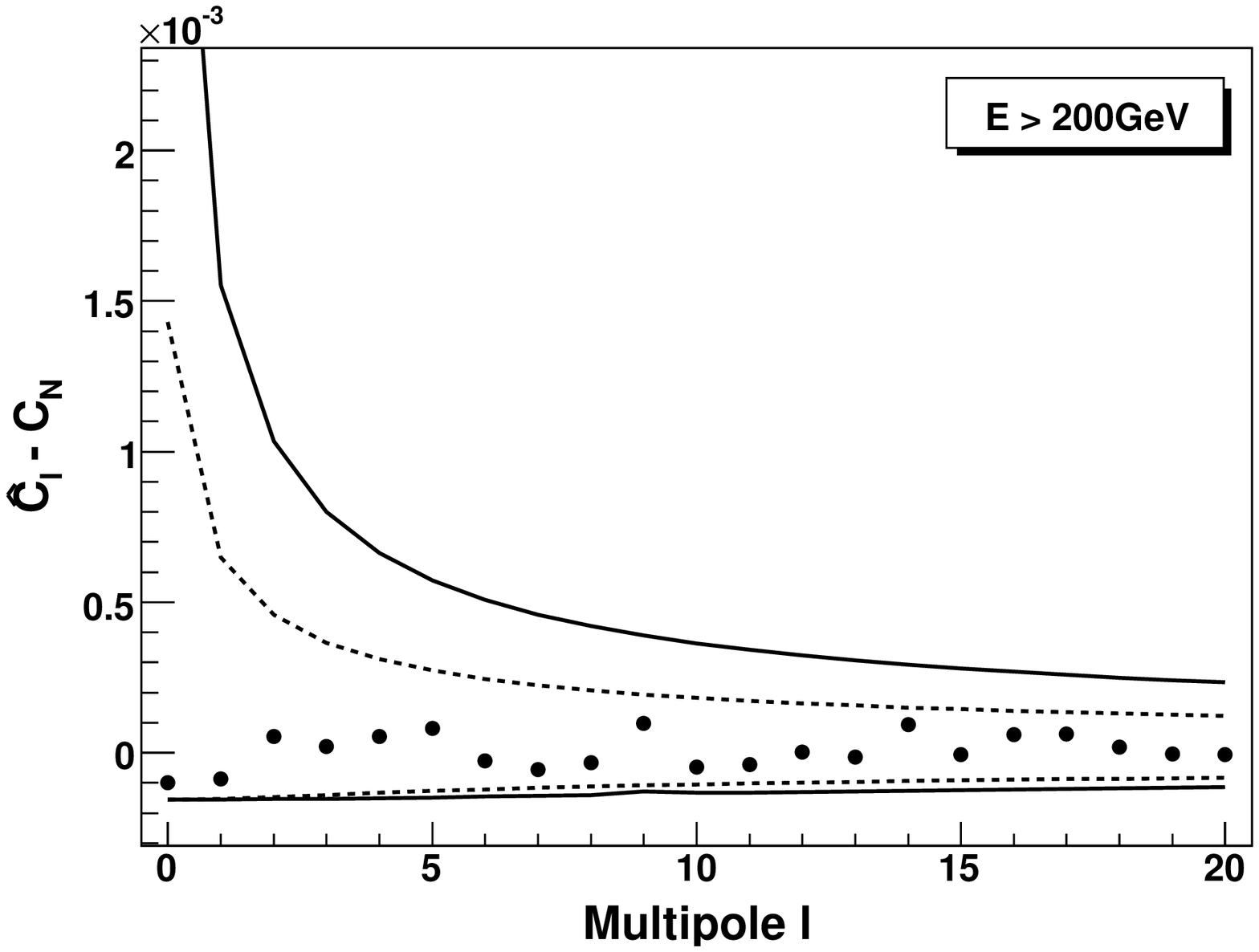}
\includegraphics[width=0.48\linewidth]{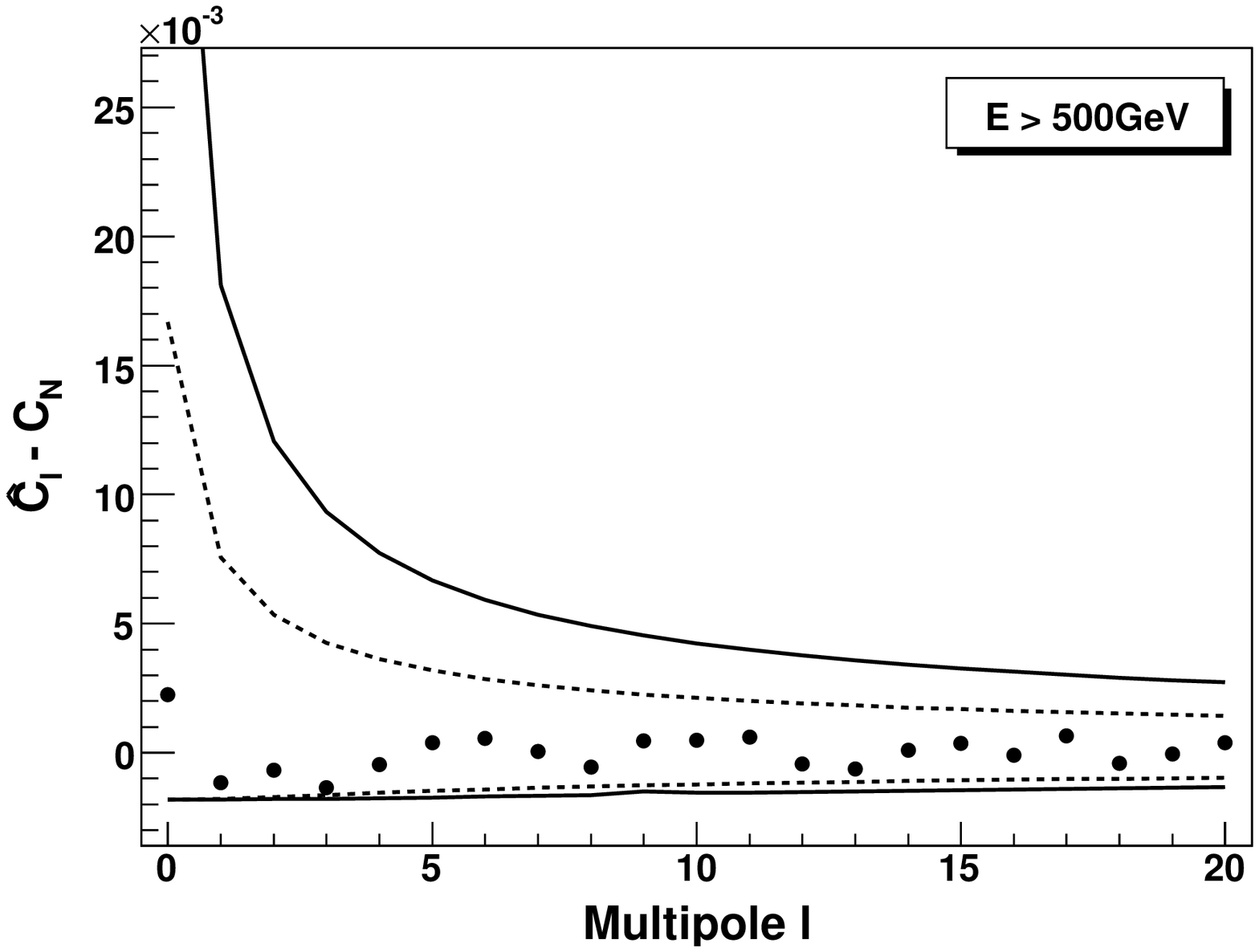}
\caption{Angular power spectra for different minimum energies: 
$60 \units{GeV}$ (top left panel), $100\units{GeV}$ (top right panel),
$200 \units{GeV}$ (bottom left panel), $500\units{GeV}$ (bottom right panel).
The points show the quantities $\hat{C}_{l} - C_{N}$. The dashed lines
and the continuous lines show respectively the $3\sigma$ and 
$5\sigma$ intervals for the probability distribution of the white
noise. 
\label{fig:powspectra}}
\end{figure*}

The previous analyses excluded variations of the CRE flux
correlated from the Sun's direction. However, due to the effects
of the heliospheric magnetic field and the geomagnetic
field, a CRE signal from the Sun could produce a flux
excess from a direction shifted with respect to the Sun's
position. Also, since CREs from the Sun are expected to
be spread over a cone with a finite angular radius, an
excess of CREs from the Sun could induce an anisotropy
on a large angular scale.

To investigate this possibility we implemented a more general
analysis method, based on spherical harmonics analysis of
a fluctuation sky map. This method was applied in
Ref.~\cite{Ackermann:2010ip} to search for anisotropies
in the CRE flux in the Galactic reference frame, 
while in this paper we adopted the custom 
coordinates in Eq.~\ref{eq:coord} derived from the ecliptic
reference frame.

The fluctuation sky map was built starting from 
the real sky map and from the simulated one, which was 
generated using the randomized data sets. The analysis was
performed on sky maps of the counts integrated above
a given energy in order to retain
sufficient statistics in each energy bin. The fluctuation 
in the $i$th pixel is:

\begin{equation}
f_{i}(>E) = \cfrac{N_{i,real}(>E) - \alpha N_{i,sim}(>E)}
{\alpha N_{i,sim}(>E)}.
\label{eq:fluctuation}
\end{equation}

The fluctuation sky map is then expanded in the basis of 
spherical harmonics to obtain the set of coefficients $a_{lm}$. 
The coefficients of the angular power spectrum 
are given by the variance of 
the $2l+1$ $a_{lm}$ coefficients at each multipole:

\begin{equation}
\hat{C}_{l} = \cfrac{1}{2l+1} \sum_{m=-l}^{l} | a_{lm} |^{2}
\end{equation} 
Each coefficient $\hat{C}_{l}$ characterizes the intensity of 
the fluctuations on an angular scale of $\sim 180\degrees /l$.
In the case of an isotropic flux, each of the coefficients
$C_{l}$ can be seen as a random variable with a true
value (white noise) given by:

\begin{equation}
C_{l} = C_{N} = \cfrac{4 \pi}{N}
\end{equation}
where $N$ is the total number of observed events.
The confidence intervals for the $\hat{C}_{l}$ can be
evaluated by noting that the random variable
$(2l+1)\hat{C}_{l}/C_{l}$ follows a $\chi^{2}_{2l+1}$
distribution.

In Fig.~\ref{fig:powspectra} the angular power spectra
after the subtraction of the white noise contribution
$C_{N}$ are shown for four different minimum energies 
($60\units{GeV}$, $100\units{GeV}$, $200\units{GeV}$ 
and $500\units{GeV}$). The curves show the $3\sigma$ 
and $5\sigma$ probability
intervals assuming the hypothesis of an isotropic
CRE flux. All the data points lie within the $3\sigma$ 
interval, indicating that the measurements are consistent
with the hypothesis of an isotropic CRE flux. 
Hence, we conclude that
no preferred CRE arrival directions are observed.

\section{Solar CRE fluxes from dark matter}

We now determine constraints on DM model parameters by 
comparing our upper limits on solar CRE fluxes to the predicted 
fluxes of the two DM annihilation scenarios considered 
in Ref.~\cite{Schuster:2009fc}: (1) capture of DM particles 
by the Sun via elastic scattering interactions and 
subsequent annihilation to $e^{\pm}$ through an 
intermediate state $\phi$, and (2) capture of DM 
particles by the Sun via inelastic scattering interactions 
and subsequent annihilation of the captured DM particles 
outside the Sun directly to $e^{\pm}$.

\subsection{Dark matter annihilation through an intermediate state}
\label{sec:intstate}

In this case we assume the standard scenario for WIMP 
capture by the Sun, namely that DM particles $\chi$ 
are captured by the Sun through elastic scattering interactions 
and then continue to lose energy through subsequent scatterings, 
eventually thermalizing and sinking to the core where they annihilate.  
In general, the only annihilation products which can escape 
the Sun are neutrinos; photons and charged particle final states 
are trapped by interactions with the dense matter in the Sun.  
However, recently scenarios have been proposed in which DM 
particles annihilate into a light intermediate state $\phi$, 
i.e., $\chi\chi \rightarrow \phi\phi$, with the $\phi$ subsequently 
decaying to standard model 
particles;
these models have been suggested to provide a means
of explaining an excess in the CRE spectrum 
reported by ATIC and \emph{Fermi} and in the positron fraction
reported by PAMELA by DM annihilation or decay 
\cite{Pospelov:2008jd,Cholis:2008wq,Cholis:2008qq,Kuhlen:2009is,Bergstrom:2009fa,
Grasso:2009ma,Cirelli:2010nh}. 
For the case considered in Ref.~\cite{Schuster:2009fc}, 
the $\phi$ are assumed to be able to escape the Sun without 
further interactions, with each $\phi$ decaying to an $e^{\pm}$ pair.  
If this decay happens outside the surface of the Sun, 
the $e^{\pm}$ could reach the Earth and may be detectable in 
the form of an observed excess of CREs from the direction 
of the Sun.  

The DM particles are assumed to annihilate at rest in the 
core of the Sun, so in the lab frame the energy of 
the $\phi$, $E_{\phi}$, is equal to the DM particle mass 
$m_{\chi}$.  We assume $\phi$ to be a light scalar such 
that $m_{\phi} \ll m_{\chi}$, hence the $\phi$ are relativistic.  
The energy of the $\phi$ is described by the parameter 
$\beta_{\rm cl} = v_{\rm cl}/c$ where $v_{\rm cl}$ is 
the relative velocity of the lab frame and the $\phi$ 
rest frame (hereafter, the CM frame).  The $\phi$ are 
assumed to lose a negligible amount of energy exiting the Sun.
A $\phi$ decays into an $e^{\pm}$ pair with an isotropic angular 
distribution in the CM frame.  Both the $e^{+}$ and $e^{-}$ have the same energy 
in the CM frame, parameterized by 
$\beta_{\rm jc} = v_{\rm jc}/c = \sqrt{1-(4 m_{e}^{2}/m_{\phi}^{2})}$, 
where $m_{e}$ is the electron mass and $v_{\rm jc}$ is 
the velocity of particle $j$ (the electron or positron) 
in the CM frame.  However, in the lab frame the $e^{\pm}$ are 
boosted and so the angular distribution is no longer isotropic, 
and the energy of an $e^{\pm}$ in the lab frame depends on 
the angle at which it is emitted relative to the direction 
in which the $\phi$ is traveling. In the lab frame the angle 
at which the $e^{\pm}$ is emitted is denoted $\theta_{\rm lab}$; 
in the CM frame it is $\theta_{\rm cm}$.  
The $e^{\pm}$ are assumed to travel in straight lines 
and to suffer no energy losses before reaching the detector, 
i.e., the effects of magnetic fields are assumed to be negligible.

The flux of $e^{\pm}$ per time, area, and energy per cosine 
of the detector angle from the direction $\theta_{\rm det}$ 
is given by integrating the differential rate of decay of 
$\phi$ along the line of sight in the direction of $\theta_{\rm det}$,
\begin{eqnarray}
\label{eq:flux}
\frac{dN}{dt\, dA\, d\cos\theta_{\rm det} dE_{\rm det}}(\theta_{\rm det},E_{\rm det}) = \\ \nonumber
\int_{0}^{\infty}\!\!\!dR\;
\frac{dN}{dV dt}\; 
\frac{d\Gamma}{d\cos\theta_{\rm det}} 
\delta(E_{\rm det} - E),
\end{eqnarray}
where $R$ is the distance from the detector in the 
line-of-sight direction defined by $\theta_{\rm det}$.  
At each $\theta_{\rm det}$ we exclude the contribution 
from $\phi$ decays occurring in the range of $R$ within the surface of the Sun.

The first term in Eq.~\ref{eq:flux} is the rate of 
production of $e^{\pm}$ (equal to twice the rate 
of $\phi$ decays) per volume at a volume element 
a distance $r$ from the center of the Sun, and is given 
by

\begin{equation}
\label{eq:decayrate}
\frac{dN}{dV dt}\left(r\left(\theta_{\rm det},R\right)\right)  = 
2\, \frac{C_{\odot}e^{-r/L}}{4 \pi r^{2} L}
\end{equation}
where $C_{\odot}$ is the capture rate of DM particles in 
the Sun and the characteristic decay length 
$L \equiv \gamma_{\rm cl} c \tau$, 
where $\gamma_{\rm cl} = \gamma(\beta_{\rm cl})$ 
with $\gamma(\beta) = 1/\sqrt{1-\beta^{2}}$ and $\tau$ 
is the lifetime of the $\phi$.  Equilibrium is assumed, 
i.e., for every two DM particles that are captured, 
two annihilate, so $C_{\odot}$ also represents the rate 
of production of $\phi$ particles by annihilation at the 
center of the Sun.  For the range of scattering cross-sections 
considered here, the capture rate is in general sufficiently large 
that equilibrium is a valid assumption, although we discuss this 
issue in greater detail when presenting our constraints.

We consider separately the cases of solar capture by 
spin-independent scattering and spin-dependent scattering. 
Using DarkSUSY~\cite{Gondolo:2004sc}, we calculate 
the capture rate via elastic scattering $C_{\odot}$ 
as a function of $m_{\chi}$.  A Maxwell-Boltzmann velocity
distribution is assumed, with the solar velocity relative
to the DM rest frame $v_{\odot} = 220 \units{km/s}$, the
DM velocity dispersion $\tilde{v} = 270 \units{km/s}$, and the local
DM density $\rho_{\rm DM} = 0.3 \units{GeV/cm^{3}}$.  For the case of spin-independent
scattering we set the spin-dependent scattering 
cross-section $\sigma_{\rm SD}=0 \units{cm^{2}}$ and calculate
the capture rates for $\sigma_{\rm SI}=10^{-43} 
\units{cm^{2}}$; for spin-dependent scattering we set
$\sigma_{\rm SI}=0 \units{cm^{2}}$ and calculate
the capture rates for $\sigma_{\rm SD}=10^{-40} 
\units{cm}^{2}$.  The reference values of $\sigma_{\rm SI}$
and $\sigma_{\rm SD}$ roughly correspond to the
current experimental upper limits on these parameters.
For the range of parameters considered here the capture rate 
scales linearly with $\sigma_{\rm SD}$ and $\sigma_{\rm SI}$.

The second term in Eq.~\ref{eq:flux} is the angular 
distribution of the $e^{\pm}$ from the decay of 
a $\phi$, as observed in the lab frame, and expressed in 
terms of detector angle,

\begin{equation}
\frac{d\Gamma}{d\cos\theta_{\rm det}} = \frac{d\Gamma}{d\cos\theta_{\rm cm}} \; \left| \frac{d\cos\theta_{\rm cm}}{d\cos\theta_{\rm lab}}\right| \; \frac{d\cos\theta_{\rm lab}}{d\cos\theta_{\rm det}}.
\end{equation}
In the rest frame of the $\phi$ the decays are isotropic, 
so, after integrating over the azimuthal angle we can write

\begin{equation}
\label{eq:angdistrib}
\frac{d\Gamma}{d\cos\theta_{\rm cm}} = -1/2.
\end{equation}
The transformation between the CM angle and lab angle is 
given by~\cite{Dick:2009}

\begin{equation}
\left| \frac{d\cos\theta_{\rm cm}}{d\cos\theta_{\rm lab}} \right|= 
\frac{[\gamma_{\rm cl}^{2} 
(\alpha + \cos\theta_{\rm cm})^2+\sin^{2} 
\theta_{\rm cm}]^{3/2}}{\left|\gamma_{\rm cl}(1 + \alpha 
\cos\theta_{\rm cm})\right|},
\end{equation}
with $\gamma_{\rm cl} = \gamma(\beta_{\rm cl})$ 
and $\alpha = \beta_{\rm cl}/\beta_{\rm jc} $. 
The lab and detector angles are related by

\begin{equation}
\theta_{\rm lab}=\theta_{\rm det} + 
 \sin^{-1}\left(\frac{R \sin\theta_{\rm det}}{r}\right),
\end{equation}
which gives

\begin{equation}
\frac{d\cos\theta_{\rm lab}}{d\cos\theta_{\rm det}} = 
\frac{(| D_{\odot} - R\cos(\theta_{\rm det}) | +
	R\cos(\theta_{\rm det}) )^{2}}
	{r | D_{\odot} - R\cos(\theta_{\rm det}) |}.
\end{equation}

The delta function in Eq.~\ref{eq:flux} enforces 
that the energy observed at the detector is equal 
to the energy of the emitted $e^{\pm}$ boosted to the lab frame,

\begin{equation}
\label{eq:energydelta}
E(\theta_{\rm cm}) = \frac{1}{2} \gamma_{\rm cl} m_{\phi } (1 + \beta_{cl} \cos\theta_{\rm cm}).
\end{equation}
Note that because the energy in the lab frame depends 
only on $\theta_{\rm cm}$, and because $\theta_{\rm lab}$ 
is determined by $\theta_{\rm cm}$, fixing $E_{\rm det}$ 
corresponds to selecting only CREs emitted at the 
corresponding $\theta_{\rm lab}$.  For a specified 
$\theta_{\rm det}$, the $\theta_{\rm lab}$ of particles 
observed along the line-of-sight $R$ varies, hence the 
observed energy of CREs emitted from a point along the 
line-of-sight is a function of $R$, i.e., $E_{\rm det}(R)$.  
We rewrite the delta function in Eq.~\ref{eq:flux} as 
the composition 

\begin{equation}
\delta(E_{\rm det} - E(R)) 
=\cfrac{\delta(R-R_{0})}{\cfrac{{\rm d}E}{{\rm d}R}(R_{0})}
\end{equation}
and then perform the integration over $R$.  
The parameter $R_{0}$ is the value of $R$ along the 
line-of-sight in the direction $\theta_{\rm det}$ 
where $\theta_{\rm lab}$ takes the value required to 
generate CREs with a given $E_{\rm det}$.

We evaluate the CRE flux within a ROI of $30 \degrees$ 
centered on the Sun, and fix the value of $m_{\phi}=1\units{GeV}$. 
We calculate limits for three values of the decay length $L=5\units{AU}$,
$1\units{AU}$, and $0.1\units{AU}$.
Decreasing $L$ increases the observed CRE flux by condensing the
region within which most $\phi$ decay.  However, we 
emphasize that even for as large a decay length as $L=5\units{AU}$, 
the signal in the energy range used in this analysis is 
strongly peaked in the direction of the Sun and extends 
only a few degrees at most.  Since the $\phi$ in this scenario 
are relativistic, in the lab frame the emitted $e^{\pm}$ are 
boosted along the direction the $\phi$ is moving, and so 
only $\phi$ exiting the Sun very close to the direction 
of the detector will produce decay products with large 
enough $\theta_{\rm lab}$ to reach the detector.  
In particular, for the $e^{\pm}$ to have sufficient energy 
to fall within the energy range of this analysis, a significant 
fraction of the $\phi$ energy must be deposited into 
the $e^{\pm}$ that reach the detector.  This only occurs 
for $e^{\pm}$ emitted with very small $\theta_{\rm lab}$.  
This also leads to an energy dependence of the angular 
signal: for a given DM scenario, the angular 
extent of the flux at high energies is smaller than 
at lower energies. We note that decreasing $m_{\phi}$ for a 
fixed $m_{\chi}$ narrows the angular extent of the signal, 
and therefore has little impact on our results.  We confirmed
that for $m_{\phi}$ as large as $10\units{GeV}$, the cross-section limits
vary negligibly except for a slight weakening of the limit at the lowest end 
of the $m_{\chi}$ range considered here.

\begin{figure}
\includegraphics[width=0.48\textwidth]{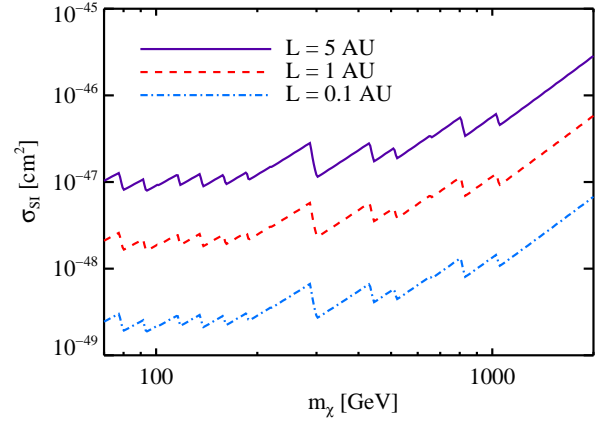}
\caption{Constraints on DM annihilation to 
$e^{+}e^{-}$ via an intermediate state, from solar CRE flux upper limits.
Solar capture of DM is assumed to take place via 
spin-independent scattering.  The constraints obtained for three values 
of the decay length $L$ of the intermediate state are shown.  
Models above the curves exceed 
the solar CRE flux upper limit at $95\%$ CL for a $30\degrees$ ROI 
centered on the Sun.  
\label{fig:intstatelimSI}}
\end{figure}

\begin{figure}
\includegraphics[width=0.48\textwidth]{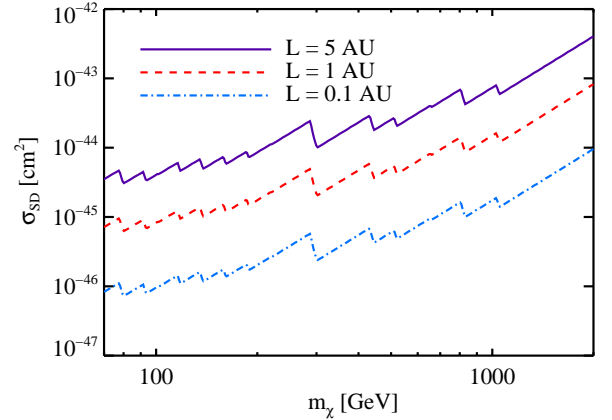}
\caption{Constraints on DM parameters for annihilation to 
$e^{+}e^{-}$ via an intermediate state as in Fig.~\ref{fig:intstatelimSI},
except assuming solar capture by spin-dependent scattering.
\label{fig:intstatelimSD}}
\end{figure}

Figs.~\ref{fig:intstatelimSI} and~\ref{fig:intstatelimSD} show the 
constraints on 
$\sigma_{\rm SI}$ and $\sigma_{\rm SD}$ as a function of $m_{\chi}$, 
derived from the upper limits on the solar CRE flux
obtained in~\S\ref{sec:isotropic}.  For each $m_{\chi}$ 
the CRE flux in each energy bin used in this analysis was 
calculated, and the limit on the scattering cross-section 
was set by the energy bin providing the strongest constraint.  
The jagged shape of the curve reflects the transitions between 
the energy bins setting the strongest limit.  
Models above the curves exceed the $95\%$ CL solar CRE flux 
upper limit for the $30\degrees$ ROI in at least one energy bin.
Ref.~\citep{Schuster:2009fc} notes that due to the Parker spiral shape
of the Sun's magnetic field, CREs emitted from the Sun may be deflected
in such a way as to appear to originate from a source displaced by 
up to $30\degrees$ from the Sun's position.  If we instead consider
larger ROIs centered on the Sun in order to accommodate the expected
angular distribution of the flux of a displaced source, the constraints derived 
on the scattering cross-sections would be weakened by $\sim 30\%$
using the flux upper limit for the $45\degrees$ ROI, 
or by a factor of $\sim2$ if the $60\degrees$ ROI flux
upper limit were used.

The bounds on the scattering cross-sections
we derive for $e^{\pm}$ final states
are significantly below the typical constraints from direct detection
experiments, and so we are prompted to examine more closely the validity
of our assumption of equilibrium.
For the limiting values we derive on elastic scattering cross-sections,
capture and annihilation are effectively in equilibrium assuming an
annihilation cross-section consistent with thermal relic dark matter
$\langle \sigma v \rangle = 3 \times 10^{-26} \units{cm^{3}} \units{s^{-1}}$
for all values of the decay length $L$
considered here.  In particular, for the limiting values of the scattering cross-sections
the flux suppression relative to the equilibrium flux for any mass we consider
is always less than 3\% (following the
standard calculation implemented in Ref.~\cite{Gondolo:2004sc}), and
thus we work under the assumption of equilibrium,
noting that there remain uncertainties in the capture rate calculation
at the level of a factor of a few (e.g., \cite{Sivertsson:2009nx}).

Decays to $e^{\pm}$ are generally accompanied by final state
radiation (FSR), so these scenarios can also be constrained
by solar gamma-ray observations.  Ref.~\cite{Schuster:2009au}
derived bounds on the rate of decay to $e^{\pm}$ by requiring
that the predicted FSR does not exceed the solar gamma-ray
emission measured by \emph{Fermi}.
However, the constraints we
obtained on the elastic scattering cross-sections from the
solar CRE flux correspond to constraints on
the annihilation rate roughly 2-4 orders of magnitude stronger than
those placed by gamma-ray constraints on FSR.  The strength of
the CRE limits relative to those from FSR increases for
larger $m_\chi$.
The relative strength of the constraints derived from the CRE
flux limits compared to those from the gamma-ray measurements
can be attributed in part to the fact that the FSR flux
produced by annihilation to $e^{+}e^{-}$
is $\sim2$-3 orders of magnitude smaller than the CRE flux.
FSR emission also must compete with a known background
gamma-ray flux from the Sun~\cite{Giglietto:2010}.  Furthermore,
the FSR constraints in Ref.~\cite{Schuster:2009au}
were derived using the preliminary \emph{Fermi}
measurement of the solar spectrum which extends only to $10 \units{GeV}$,
while this analysis spans CRE energies from 
$60 \units{GeV}$ to $\sim 1 \units{TeV}$.
Since the FSR photon spectrum is harder than the measured solar
gamma-ray spectrum, the strongest constraints are obtained 
from the highest energy bin in that analysis.  Due to the 
fact that the FSR spectrum associated with the DM mass range considered
in this analysis extends substantially higher 
than $10 \units{GeV}$, the existing FSR constraints
are significantly less competitive than our CRE constraints.  
A measurement of the solar gamma-ray
emission at higher energies could likely strengthen 
the FSR constraints to some extent.

\subsection{Inelastic dark matter}

We now consider the flux of $e^{\pm}$ from annihilation of 
DM particles captured by the Sun but with orbits which take them 
outside the surface of the Sun. In a standard WIMP scenario, 
DM particles captured by the Sun via elastic scattering quickly 
undergo subsequent scatterings which cause them to settle to the 
core, and hence the fraction of captured DM particles outside the 
surface of the Sun at any given time is negligible \cite{Sivertsson:2009nx}.  
However, this is not necessarily the case for inelastic 
dark matter (iDM)~\cite{TuckerSmith:2001hy,
Finkbeiner:2007kk,ArkaniHamed:2008qn}.  This class of models 
has garnered interest recently in light of claims that 
iDM could naturally explain  such observations as the 
$511 \units{keV}$ line observed by INTEGRAL/SPI~\cite{Finkbeiner:2007kk}
and the apparently inconsistent results of DAMA/LIBRA and CDMS 
if the DM scattered inelastically and thereby 
transitioned to an excited 
state with a slightly heavier mass.

For a DM particle $\chi$ to scatter inelastically off a nucleon 
$N$ via the process $\chi + N \rightarrow \chi^{\star} + N$, 
the DM must have energy $E \ge \delta(1+m_{\chi}/m_{N})$, 
where $\delta=m_{\chi^{\star}}-m_{\chi}$.
Particles captured by the sun by inelastic scattering typically 
lose enough energy after only a few interactions to prevent 
further energy loss by scattering.  If the elastic scattering 
cross-section is sufficiently small 
($\sigma_{\rm n} \lesssim 10^{-47} \units{cm^{2}}$, e.g., 
Ref.~\cite{Schuster:2009fc}), the captured particles will be 
unable to thermalize and settle to the core, and instead 
will remain on relatively large orbits.  As a result, the 
density of captured DM particles outside the Sun may not be 
negligible in an iDM scenario, and the annihilation of 
those particles to $e^{\pm}$ could thus produce an observable flux of 
CREs from the direction of the Sun.  While it is not necessary
for DM to annihilate primarily to $e^{\pm}$ in order to explain the
direct detection results (since direct detection experiments are not
sensitive to the dominant annihilation channels), leptophilic
iDM is strongly motivated since it could provide a consistent
interpretation of multiple data sets \cite{Finkbeiner:2007kk,
ArkaniHamed:2008qn,Batell:2009zp,Cholis:2009va}.

In the following we will assume that the DM particles annihilate 
at rest and thus the energy of the $e^{\pm}$ produced in annihilation 
is well-approximated by $E_{\rm CRE} = m_{\chi}$.  
We will further assume that the CREs suffer no significant 
energy losses between production at the surface of the Sun and 
arrival at the detector, and so we expect a mono-energetic 
flux of CREs in this scenario.  

For simplicity, we assume all annihilations occur at the 
surface of the Sun (as in \cite{Schuster:2009fc}), since the density 
of DM falls off quickly with distance from the Sun.  
Naturally, $e^{\pm}$ produced in annihilations inside the 
surface of the Sun cannot escape the Sun, and thus do not 
produce a detectable flux.  

The isotropic flux of $e^{\pm}$ particles from the Sun is
\begin{equation}
\label{eq:idmfluxeqn}
F = 2\, \frac{\Gamma_{\rm A, out}}{4 \pi D_{\odot}^{2}}
\end{equation}
where $\Gamma_{\rm A, out}$ is the annihilation rate 
of DM particles outside the surface of the Sun.  The 
factor of 2 accounts for the fact that 2 CREs are emitted 
per annihilation of a pair of DM particles.  However, it 
is also necessary to take into account that CREs produced 
on the surface of the Sun opposite to the Earth are 
extremely unlikely to reach the detector, so we assume 
the flux of CREs \emph{observable} at the detector is 
a factor of 2 smaller than that given by Eq.~\ref{eq:idmfluxeqn}.

Following Refs.~\citep{Nussinov:2009ft, Menon:2009qj}, 
we assume that capture and annihilation of particles in this 
scenario is in equilibrium, i.e., 
$\Gamma_{\rm A} = \frac{1}{2}C_{\odot}$, 
where $\Gamma_{\rm A}$ is the total annihilation 
rate at all radii.  We emphasize, however,
that due to significant uncertainties in the
density profile of the captured iDM particles,
the assumption of equilibrium is less robust in this case
than in the elastic scattering scenario.  
Ref.~\cite{Nussinov:2009ft}
concludes that equilibrium will be attained, but notes the sizable
uncertainties in this calculation.  
On the other hand, for the limiting cross-sections
we determine for this scenario,
the condition for equilibrium given in Ref.~\cite{Menon:2009qj}
for inelastic capture
requires a minimum annihilation cross-section ranging from more than an order of magnitude
smaller than for a thermal relic for small masses and $\delta=110\units{keV}$
to a factor of $\sim 3$ larger than thermal for larger masses and $\delta=140\units{keV}$.
In light of the uncertainties in this calculation, we again work under the assumption of
equilibrium
when deriving limits on the scattering cross-section.

Defining $f_{\rm out}$ as the 
fraction of captured DM particles outside 
the Sun at a given instant, we have
\begin{equation}
\Gamma_{\rm A, out} = 
f_{\rm out} \Gamma_{\rm A} = \frac{1}{2}f_{\rm out} C_{\odot}.
\end{equation}
The capture rate of iDM particles by the Sun $C_{\odot}$ 
was calculated by Refs.~\citep{Nussinov:2009ft, Menon:2009qj}.  
Both studies note that there are uncertainties in this 
calculation at the factor of a few level.  We use the 
capture rate as a function of DM mass $m_{\chi}$ 
and mass splitting $\delta$ as given in Fig.~2 of 
Ref.~\citep{Menon:2009qj}, and interpolate the results 
shown in that figure. The capture rates were calculated 
assuming the following parameters: the velocity of the 
Sun in the DM rest frame $v_{\odot} = 250 \units{km/s}$, 
the DM velocity dispersion $\tilde{v} = 250 \units{km/s}$, 
the local DM density $\rho_{\rm DM} = 0.3 \units{GeV/cm^{3}}$, 
and the cross-section per nucleon in the elastic limit 
$\sigma_{0} = 10^{-40} \units{cm^{2}}$.  The relation 
between the total inelastic scattering cross-section 
and the total elastic scattering cross-section is 
given in Eq.~7 of Ref.~\citep{Menon:2009qj}.  
The capture rate scales linearly with $\rho_{\rm DM}$ 
and $\sigma_{0}$, while the dependence 
on $v_{\odot}$ and $\tilde{v}$ is mild over the 
mass range of interest ($m_{\chi} \sim 100 \units{GeV}$ 
to $\sim 1 \units{TeV}$).  We note, however,
that the constraints obtained 
by direct detection experiments
may be more sensitive to variations in the assumed
velocity distribution of the DM particles.

The parameter $f_{\rm out}$ was calculated by 
Ref.~\citep{Schuster:2009fc} by simulating the capture of 
DM particles by the Sun via inelastic scattering.  
Here we interpolate the values of $f_{\rm out}$ as 
a function of $\delta$ shown in Fig.~4 of that work, 
which were calculated for $m_{\chi} = 1 \units{TeV}$.  
Those authors note that the dependence on $m_{\chi}$ is weak 
for the mass range of interest, thus we adopt the values 
of $f_{\rm out}$ determined by \cite{Schuster:2009fc} 
for $m_{\chi} = 1\units{TeV}$ for all masses considered.
We caution that the calculation of $f_{\rm out}$
is subject to severe uncertainties, and a detailed study
beyond the scope of this work
is needed to more robustly estimate the value of this parameter.
In particular, we note that $f_{\rm out}$ varies
by more than an order of magnitude over the range of $\delta$
values considered in this study,
and we therefore stress that the calculation of $f_{\rm out}$
introduces uncertainties in the derived scattering cross-section limits 
of at least a factor of a few.

We calculate the flux of CREs from annihilation of 
DM in this scenario as a function of $m_{\chi}$ and 
$\sigma_{0}$ for three values of the parameter $\delta$.  
We then derive constraints on the $m_{\chi}$-$\sigma_{0}$ 
parameter space by requiring that the predicted flux of 
each DM model does not exceed the $95\%$ CL upper limits 
on solar CRE fluxes for a $30\degrees$ ROI centered on the 
Sun, again using the results derived in~\S\ref{sec:isotropic}.  
Since the region from which the DM-induced flux 
originates in this scenario is the angular extent of 
the Sun, the $30\degrees$ ROI is more than sufficient 
to encompass all of the DM signal. 

\begin{figure}
\includegraphics[width=0.48\textwidth]{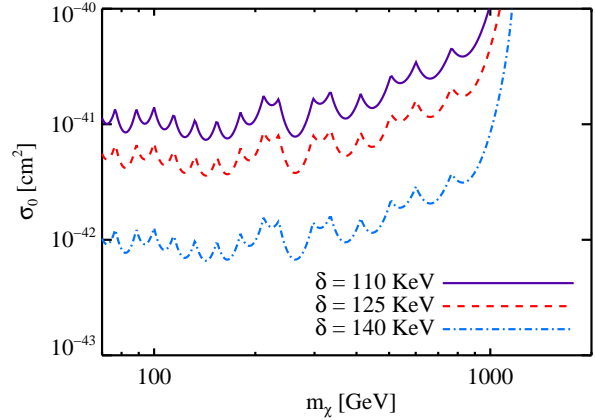}
\caption{Constraints on iDM model parameters for three values 
of the mass splitting $\delta$. Models above the curves produce 
a solar CRE flux that exceeds the $95\%$ CL flux upper limit 
for a $30\degrees$ ROI centered on the Sun in one or more energy bins. 
\label{fig:idmlim}}
\end{figure}

The predicted flux is mono-energetic, however the finite 
energy resolution of the LAT will result in the observed 
events being assigned to more than one energy bin.  
Since this may have a non-negligible impact on the derived 
scattering cross-section limits for DM masses near the 
energy bin edges, we convolve the predicted signal from 
each model with the energy resolution of the LAT and 
calculate its flux in each energy bin used in the analysis.  
We approximate the energy dispersion of the LAT as a Gaussian 
with $\sigma$ given by the half-width of the $68\%$ containment 
window (see Fig.~9 of \cite{Ackermann:2010ij}).  
For the energy range considered here the energy resolution 
ranges from $\sim 5\%$ to $\sim 14\%$.  The cross-section 
limit at each mass is obtained from the energy bin providing 
the strongest constraint.  

Fig.~\ref{fig:idmlim} shows the constraints from the solar 
CRE flux upper limits on iDM models in the 
$m_{\chi}$-$\sigma_{0}$ parameter space for three values 
of $\delta$.  Models in the regions above the curves exceed 
the $95\%$ CL flux upper limit for the $30\degrees$ ROI in 
at least one energy bin.  The rounded shape of the curves 
is due to accounting for the energy resolution of the LAT.  
These limits exclude the regions of parameter space compatible 
with the results of DAMA/LIBRA and CDMS (in addition to several
other direct detection experiments) as determined 
by Ref.~\cite{Ahmed:2010hw} for $\delta=120\units{keV}$, for the range of masses 
accessible to our analysis ($m_\chi \gtrsim 70\units{GeV}$),
assuming the dominant annihilation channel is
$e^{\pm}$.  Models consistent with both DAMA/LIBRA and CDMS at 90\% CL
exist for values of $\delta$ ranging from $\sim 85\units{keV}$ to 
$\sim 135\units{keV}$~\cite{Ahmed:2010hw};
for masses from 70~GeV to 250 GeV the range of 
allowed scattering cross-sections is from $\sigma_{0} \sim 10^{-40} \units{cm^{2}}$
to $\sigma_{0} \sim 10^{-39}$ cm$^{2}$~\cite{Chang:2008gd}.
Although the uncertainties in the calculation of the 
DM fluxes in this scenario are significant, we emphasize that
constraining $\sigma_{0} \lesssim 10^{-40}\units{cm^{2}}$ is sufficient
to exclude the cross-sections of models consistent with
both data sets.  The bounds we derive exclude
the relevant cross-sections by 1-2 orders of magnitude,
and hence we conclude that the parameter space of models
preferred by DAMA/LIBRA can be confidently ruled out for
$m_\chi \gtrsim 70\units{GeV}$ for annihilation
to $e^{\pm}$ despite the uncertainties in the flux
calculation.  

This analysis constrains DM models in which the
primary annihilation channel is to $e^{\pm}$.  We
emphasize that although other annihilation channels can
be probed by gamma-ray \cite{Atkins:2004qr,Batell:2009zp,Schuster:2009au} 
or neutrino \cite{Nussinov:2009ft,Menon:2009qj,Schuster:2009au}
measurements, the upper limits on solar CRE fluxes provide
a uniquely strong constraint on the $e^{\pm}$ final state,
which is inaccessible to neutrino telescopes since
no neutrinos are produced for this annihilation channel.

\section{Conclusions}

We used a sample of about $1.3 \times 10^{6}$ CRE
events with energies above $60\units{GeV}$ 
detected by the {\em Fermi} LAT during its first
year of data-taking to search for  flux
excesses or deficits correlated with the Sun's direction.
Two analysis approaches were implemented, and 
neither yielded evidence of an enhancement in the
CRE flux from the direction of the Sun. This result agrees 
with the more general one shown in 
Ref.~\cite{Ackermann:2010ip}, where
no evidence of anisotropies was found in CRE arrival directions above $60\units{GeV}$
in the Galactic reference frame.

We derived limits on DM models which generate a
CRE flux from the Sun's direction for the two scenarios
discussed in Ref.~\cite{Schuster:2009fc}.
In the case of annihilation of DM through an intermediate state 
and subsequent
decay to $e^{\pm}$, the upper limits on solar CRE
fluxes provide significantly stronger constraints on
the DM scattering cross-section than limits previously derived
by constraining the FSR emission associated with
this decay channel using solar gamma-ray measurements.
For the iDM scenario, the solar CRE flux upper limits
exclude the range of models which can reconcile
the data from DAMA/LIBRA and CDMS for 
$m_{\chi} \gtrsim 70\units{GeV}$, assuming DM
annihilates predominantly to $e^{\pm}$.  Since
direct detection experiments are not sensitive to
the dominant annihilation channels of the DM particles, 
other data, e.g.,
solar gamma-ray measurements and 
neutrino searches,
may be able to further constrain
these models by excluding regions of parameter
space for alternative annihilation channels. 

\begin{acknowledgments}

The {\em Fermi} LAT Collaboration acknowledges generous ongoing support
from a number of agencies and institutes that have supported both the
development and the operation of the LAT as well as scientific data analysis.
These include the National Aeronautics and Space Administration and the
Department of Energy in the United States, the Commissariat \`a l'Energie Atomique
and the Centre National de la Recherche Scientifique / Institut National de Physique
Nucl\'eaire et de Physique des Particules in France, the Agenzia Spaziale Italiana
and the Istituto Nazionale di Fisica Nucleare in Italy, the Ministry of Education,
Culture, Sports, Science and Technology (MEXT), High Energy Accelerator Research
Organization (KEK) and Japan Aerospace Exploration Agency (JAXA) in Japan, and
the K.~A.~Wallenberg Foundation, the Swedish Research Council and the
Swedish National Space Board in Sweden.

Additional support for science analysis during the operations phase is gratefully
acknowledged from the Istituto Nazionale di Astrofisica in Italy and
the Centre National d'\'Etudes Spatiales in France.

The authors thank Joakim Edsj\"{o} for his valuable contribution during the
preparation of this manuscript.

JSG thanks J.~Beacom, B.~Dasgupta, S.~Horiuchi, D.~Malyshev, and I.~Yavin for helpful discussions.

\end{acknowledgments}

\end{document}